\documentclass[aps,prc,twocolumn,showpacs,preprintnumbers,amsmath,amssymb]{revtex4}
\usepackage[dvips]{graphicx} \usepackage{eucal} \usepackage{bm}
\usepackage{float} \newcommand{\ssT}{{\scriptscriptstyle T}}

\begin{document}
\title{
 Mean-field effects on collective flow in high-energy \\
 heavy-ion collisions at 2-158$A$ GeV energies
}

\medskip

\author{M. Isse}
\author{A. Ohnishi}
\affiliation{
 Division of Physics,
 Graduate School of Science, Hokkaido University,
Sapporo, Hokkaido 060-0810, Japan }
\author{N. Otuka}
\affiliation{
Nuclear Data Center,
Department of Nuclear Energy System,
Japan Atomic Energy Research Institute,
Tokai, Ibaraki 319-1195, Japan}
\author{P. K. Sahu}
\affiliation{
Institute of Physics, Sachivalaya Marg,
Bhubaneswar 751 005, India
}
\author{Y. Nara}
\affiliation{
Institut f\"ur Theoretische Physik,
Johann Wolfgang Goethe-Universit\"at,
 Robert-Mayer-Str. 10
60325 Frankfurt am Main, Germany
}

\begin{abstract}
Collective flows in heavy-ion collisions
from AGS ($(2-11)A$ GeV) to SPS ($(40,158)A$ GeV) energies
are investigated in a nonequilibrium transport model
with nuclear mean-field (MF).
Sideward $\langle p_x \rangle$,
directed $v_1$, and elliptic flows $v_2$
are systematically studied
with different assumptions on the nuclear equation of state (EOS).
We find that momentum dependence in the nuclear MF is important
for the understanding of the proton collective flows
at AGS and SPS energies.
Calculated results with momentum dependent MF
qualitatively reproduce the experimental data
of proton sideward, directed, and elliptic flows
in a incident energy range of $(2-158)A$ GeV.
\end{abstract}

\maketitle

\section{Introduction}
\indent

Determining the nuclear equation of state (EOS) under various conditions
has been one of the largest motivations of heavy-ion physics
in these decades~\cite{scheid74,ks81,st82,gu84,bertsch,gale,wp88,cf92,
qmdmsu,da98,da00,da02,
E895-02,eosch,ki97,do86,fopi97,fopi03,E877,E895-99,E895-00,E895-02b,
NA49-98,NA49-99,NA49-03,WIG,sahu2,sc02,fopid04,msu,persram,Giessen,Maruyama1993,
stoecker2004}.
At around the saturation density, 
EOS gives the bulk properties of nuclei
such as the binding energy and the radius.
While the first principle simulations of lattice QCD are possible
for hot baryon-free nuclear matter
and matter at small baryon density
can be studied by expanding in the power series of
baryon chemical potential $\mu$~\cite{lattice},
properties of highly compressed matter are still under debate.
Thus phenomenological studies are necessary
to connect the experimental heavy-ion collision data
with the EOS especially for nuclear matter at high baryon densities.
In high-energy heavy-ion collisions,
where nuclear matter in a wide range of temperatures and densities
are probed, many ideas on the EOS and phases have been examined.
For example,
very dense matter is created
in recent RHIC experiments~\cite{qm2002}
suggesting the creation of
the gas of deconfined quarks and gluons (QGP).
In 1970's and 1980's,
the existence of strong collective flow in heavy-ion collisions
was suggested in hydrodynamics~\cite{scheid74,ks81,st82},
and it was examined in experiments at Bevalac~\cite{gu84}.
Since collective sideward flows are generated in the early stages of
collisions by the repulsive nucleon potential in nuclear matter,
the observed strong collective flows were believed to signal
very large pressure at high baryon densities,
i.e. hard EOS~\cite{bertsch}.
On the other hand,
the real part of the nucleon-nucleus potential is already repulsive
at the normal density at high incident energies,
and
the role of the momentum dependence of nuclear potential
on the collective flows was extensively studied
from around 1990~\cite{bertsch,gale,wp88,cf92,qmdmsu}.
In order to distinguish the momentum and density dependence,
we need to invoke heavy-ion collision data in a wide incident energy range.
We have now systematic collective flow data at various incident energies; 
LBNL Bevalac~\cite{eosch,ki97,do86}, 
GSI Schwerionen Synchrotron
(SIS)~\cite{fopi97,fopi03,fopid04},
MSU NSCL~\cite{msu},
BNL Alternating Gradient Synchrotron 
(AGS)~\cite{E877,E895-99,E895-00,E895-02,E895-02b},
CERN Super Proton Synchrotron
(SPS)~\cite{NA49-98,NA49-99,NA49-03},
and BNL RHIC\cite{rhic}.

Collective flow data obtained at AGS energies ($(2-11)A$ GeV)
show a good landmark to determine EOS.
As demonstrated in Ref.~\cite{sahu2},
the saturating momentum dependence of the mean-field (MF),
a large number of hadronic resonances,
and string degrees of freedom
are essential in order to explain
all of the radial, sideward, and elliptic flows at AGS energies.
The momentum dependence of nuclear potentials
in the context of collective flow was also discussed 
in Ref.~\cite{qmdmsu} with Quantum Molecular Dynamics (QMD) model.
It is discussed that we can separate the momentum dependence by
analyzing the so-called balance energy at which the flow disappears,
and it was later on confirmed in experiment at NSCL~\cite{msu}
for $E_\mathrm{inc}= (55-155)A$ MeV.
Recently, Danielewicz et al. have also discussed the EOS with these data 
within Boltzmann equation simulation~\cite{da98,da00,da02,E895-02},
showing that reliable stiffness value ($K=167-380$ MeV)
cannot be uniquely determined
from currently available collective flow data ($F$ or $v_2$)
up to AGS energies
($E_\mathrm{inc}=(0.15-11)A$ GeV)~\cite{da02}.
On the other hand,
a comparable description to theirs was obtained also in 
Relativistic Boltzmann-Uehling-Uhlenbeck ({\sc rbuu})~\cite{sahu2} 
by using a Relativistic Mean Field.
In {\sc rbuu}, MF is fitted to reproduce
the real part of the global optical potential
in Dirac phenomenology~\cite{Hama1990}.
In that work, a common MF giving $K\sim 300$ MeV is applied in the energy 
range of $(0.25-11)A$ GeV.
Thus these two works do not necessarily provide the same conclusion for
the stiffness. 
In addition, we still have large ambiguities in the MF for 
hadrons other than nucleons.
In order to reduce these ambiguities and to pin down the EOS more precisely,
recently measured flow data at lower SPS energies ($(20-80)A$ GeV)
may be helpful, because higher baryon density would be reached
at these incident energies.

Several hadronic transport models, such as
{\sc rqmd}~\cite{sorge,sorge2,Sorge1997-99,maru,RQMD_S}, {\sc bem}~\cite{da98,da00,da02},
{\sc rbuu}~\cite{sahu2,Maruyama1993,sc02}, {\sc arc}~\cite{arc},
{\sc art}~\cite{art}, {\sc hsd}~\cite{hsd},
{\sc u{\rm r}qmd}~\cite{urqmd,urqmd2} and {\sc jam}~\cite{jam}, 
have been successfully applied to describe many aspects of 
high energy heavy-ion collisions in a wide range of incident energies.
%%%%%%%%%%%
Transport models without MF effects ({\sc arc, hsd, jam})
can describe bulk observables such as transverse mass spectra
or rapidity distributions,
but they cannot explain anisotropic collective flows,
which are sensitive to MF potentials.
Transport models with MF effects ({\sc rqmd, bem, rbuu, art, u{\rm r}qmd})
have been successful in explaining
anisotropic collective flows in addition to bulk observables
up to AGS energies.
For SPS energies, however,
the MF effects on collective flows have not been
seriously investigated yet.

In this work, 
we investigate collective flows from 2$A$ GeV to 158$A$ GeV
by using a hadronic cascade model,
Jet AA Microscopic Transportation Model ({\sc jam})~\cite{jam},
combined with a covariant prescription
of MF (RQMD/S)~\cite{RQMD_S}.

This paper is organized as follows.
In Sec.~\ref{sec:eos}, we explain
our transport model and
parameterization of our EOS
used their.
In Sec.~\ref{sec:results}, we present our results
on flows on rapidity and transverse distributions
as well as their excitation functions.
In Sec.~\ref{sec:discussion},
we discuss some uncertainties in our model.
In Sec.~\ref{sec:summary},
we summarize our work.

\section{Nonequilibrium Transport Model and the Equations of State}
\label{sec:eos}
\indent

Heavy-ion collision is a dynamical process of a system
in which the temperature and density are not uniform
and the equilibrium is not necessarily reached.
Therefore, we need dynamical models to describe collisions
in order to extract static properties of nuclear matter under equilibrium.
Hydrodynamical description is the most direct way to connect 
the EOS and dynamics. 
Actually, ideal hydrodynamics has succeeded
to describe elliptic flow at low-$p_\ssT$,
up to semicentral and around mid-rapidity
at RHIC~\cite{hydroRHIC},
(However see Ref.~\cite{HiranoGyulassy} for recent reinterpretation
of the RHIC data.)
where the number of produced particles is so large that local equilibrium
may be easily achieved.
However the condition of local equilibrium may not be satisfied
up to SPS energies,
and non-equilibrium dynamics is required to study
the EOS of dense nuclear matter through heavy-ion collisions.

%%%%%%%%%%%%%%%%%%%%%%%%%%%%%%%%%%%%%%%%%%%%%%%%%%%%%%%
% Cascade
%%%%%%%%%%%%%%%%%%%%%%%%%%%%%%%%%%%%%%%%%%%%%%%%%%%%%%%
Hadron-string cascade processes are the main source of thermalization
and particle production up to SPS energies.
In the increase of incident energy from AGS
($(2-11)A$ GeV) to SPS ($(20-158)A$ GeV),
the main particle production mechanism in hadron-hadron collisions
evolves from resonance productions to string formations.
At higher energies, hard partonic interaction (jet production) becomes
more important, and the jet production cross section reaches around
20 \% of the total cross section of $pp$ at RHIC~\cite{hijing}.

{\sc jam} includes all of the above particle and jet production mechanisms,
and the applicable incident energy range is expected to be enough
(for the study of collective flow, jet production does not matter).
Inelastic hadron-hadron collisions produce resonances at low energies.
We explicitly include all established hadronic states
with masses up to around 2 GeV with explicit isospin states
as well as their antiparticles, which are made to propagate in space-time.
At higher energies
       ($\sqrt{s}\gtrsim 4$ GeV in $BB$ collisions,
       $\sqrt{s}\gtrsim 3$ GeV in $MB$ collisions,
       and $\sqrt{s}\gtrsim 2$ GeV in $MM$ collisions),
color strings are formed and they decay into hadrons
after their formation time $(\tau \sim 1$ fm/$c$)
according to the Lund string model {\sc pythia}~\cite{Sjostrand}.
%
%Hadrons
Leading hadrons having constituent quarks
can scatter within their formation time
with other hadrons assuming the additive quark cross section
which is known to be important at SPS energies~\cite{urqmd2}.
%This simulates string-hadron collisions which is known to be important
%at SPS energies~\cite{urqmd2}.

\begin{table*}
\begin{ruledtabular}
\caption{
Parameter set of density-dependent and
 momentum-dependent/independent potential.
Momentum-dependent hard (MH) and soft (MS) potential
are taken from Ref.~\cite{maru2} with simplification (See text for detail).
Momentum-independent
hard (H) and soft (S) potential are taken form
Ref.~\cite{aich}.
}
\begin{tabular}{ccccccccc}
Type &$\alpha$  &$\beta$ & $\gamma$ & $C_{\rm ex}^{(1)}$ 
 &$C_{\rm ex}^{(2)}$  &$\mu_1$ &
$\mu_2$ & $K$  \\
 & (MeV) &(MeV) & & (MeV)& (MeV) &(fm$^{-1}$) &
 (fm$^{-1}$) & (MeV) \\
\hline
MH 
&-33&110&5/3&-277&663&2.35&0.4&448\\
MS
&-268&345&7/6&-277&663&2.35&0.4&314\\
H 
& -124 & 70.5 & 2 &---&---&---&---& 380\\
S
 & -356 & 303 & 7/6&---&---&---&---& 200\\
\end{tabular}
\label{parsets}
\end{ruledtabular}
\end{table*}

\begin{figure*}[thb]
\includegraphics[width=9cm]{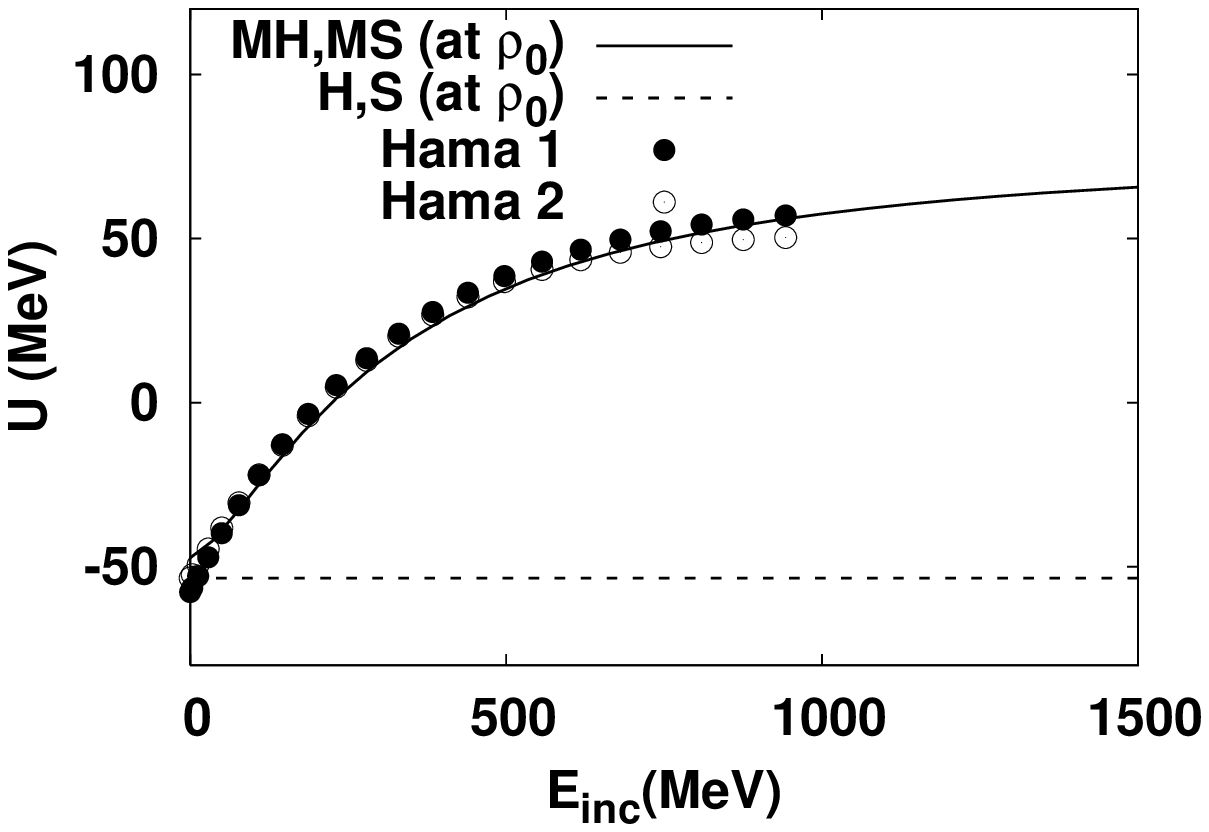}~%
\includegraphics[width=9cm]{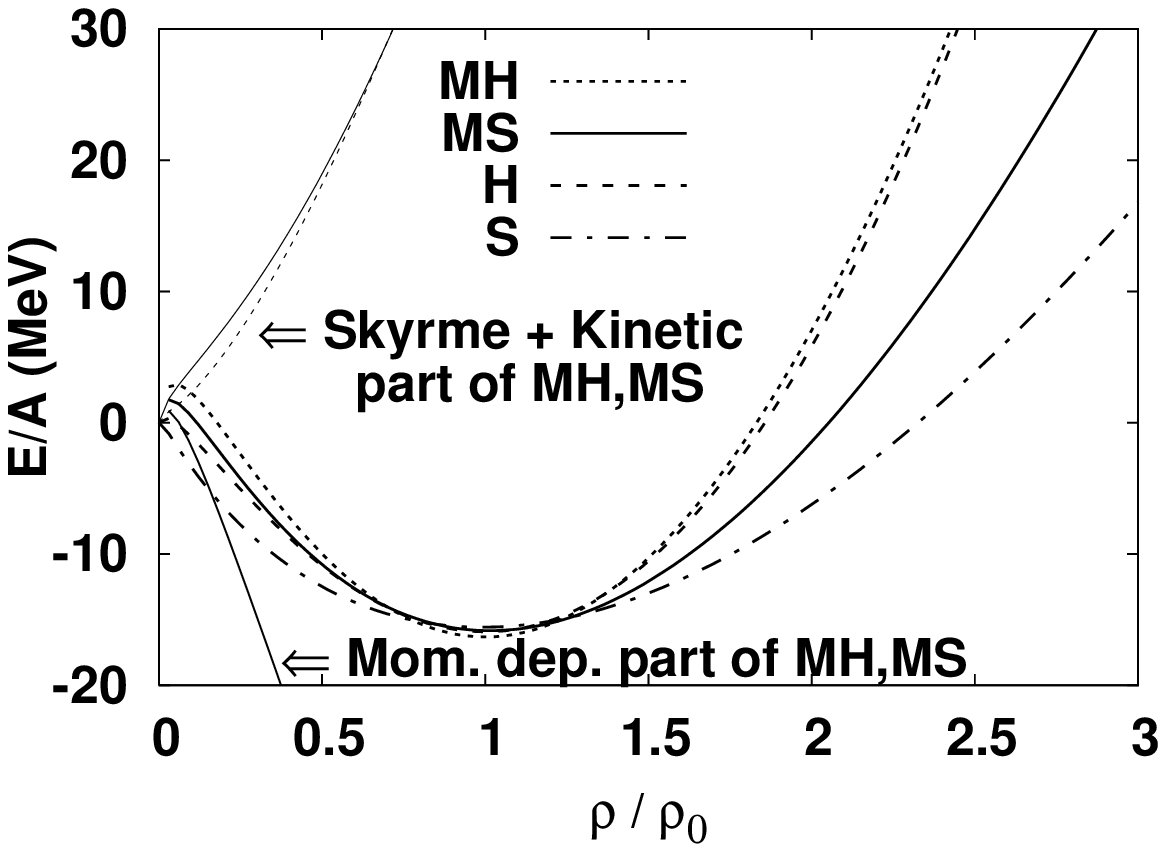}
\caption{Left:
Momentum dependence of the
single particle potentials Eq.~(\ref{umomint})
for momentum-dependent hard (MH), soft (MS)
as well as momentum-independent hard (H) and soft (S) are
compared with the real part of the global Dirac optical potential~\cite{Hama1990}.
Right:
Density dependence of total energy per nucleon
in Eq.~(\ref{eq:tote}) for 
 momentum-dependent(MH,MS) and independent(H,S) potential.
}
\label{skyeps}
\end{figure*}

%%%%%%%%%%%%%%%%%%%%%%%%%%%%%%%%%%%%%%%%%%%%%%%%%%%%%%%
% Mean Field
%%%%%%%%%%%%%%%%%%%%%%%%%%%%%%%%%%%%%%%%%%%%%%%%%%%%%%%

It is necessary to include MF effects to explain collective flow data,
and the MF should have the momentum dependence 
as well as the density dependence
in order to describe flows in a wide incident energy range.
We adopt here
a simple Skyrme type density dependent MF
in the zero-range approximation,
and a Lorentzian type momentum dependent MF~\cite{gale} 
which simulates the exchange term (Fock term)
of the Yukawa potential.
Single particle potential $U$ then has the form
\begin{eqnarray}
U(\bm r,\bm p)=&&
	\alpha\left({\rho(\bm r)\over\rho_0}\right)
	+\beta\left({\rho(\bm r)\over\rho_0}\right)^\gamma\nonumber\\
&&\hspace{-0.5cm}+\sum_{k=1,2} {C_{\rm ex}^{(k)} \over \rho_0}\!\!
\int\!\! d \bm p' {f(\bm r,\bm p') \over 1+[(\bm p - \bm p')/\mu_k]^2}
\ .
\label{usep}
\end{eqnarray}
This MF potential leads to the following total potential energy,
through a relation of $U=\delta V / \delta f$,
\begin{eqnarray}
V
	=&& \int d\bm r \biggl[
		\frac{\alpha\rho^2(\bm r)}{2\rho_0}
		+ \frac{\beta\rho^{\gamma +1}(\bm r)}{(1+\gamma)\rho_0^\gamma}
	  \biggr]\nonumber\\
	&&\hspace{-0.5cm} + \sum_{k=1,2} \frac{C^{(k)}_{\rm ex}}{2\rho_0}\!\!
		\int\!\! d\bm r d\bm p d\bm p'
		\frac{ f(\bm r, \bm p)f(\bm r, \bm p')}
		{1+[(\bm p-\bm p')/\mu_k]^2}\ ,
\label{vrho}
\end{eqnarray}
where $f(\bm r, \bm p)$ is the phase space distribution function
whose integral over $\bm p$ is normalized to the density $\rho(\bm r)$.
At zero temperature, the phase space distribution function is given as
\begin{equation}
 f(\bm r, \bm p)=\biggl(\frac{4}{3}\pi p_F^3\biggr)^{-1}
\!\!\!
\rho(\bm r)\; \Theta(p_F - |\bm p|)\label{fermigas}\ .
\end{equation} 
\begin{widetext}
Then the total energy per nucleon is
\begin{equation}
 \frac{E}{A}(\rho)  =  \frac{3}{5}\frac{p_F(\rho)^2}{2m}
  + \frac{\alpha}{2\rho_0}\rho
  + \frac{\beta}{(1+\gamma)\rho_0^\gamma}\rho^\gamma
  + \frac{\rho}{2\rho_0}
    \biggl(\frac{4}{3}\pi p_F^3 \biggr)^{-2}\!\!\!
\int^{p_F}_0 \!\!\!\!d \bm p \int^{p_F}_0\!\!\!\! d \bm p'
\!\!\sum_{k=1,2}\frac{C_{\rm ex}^{(k)}}{1+[(\bm p - \bm p')/\mu_k]^2}\ ,
\label{eq:tote}
\end{equation}
where Fermi momentum is taken to be
$ p_F(\rho)=\hbar (3\pi^2\rho/2)^{1/3}$.
See (\ref{Eq:rhoi}) for the definition of $\rho$
in the actual simulations.
Integrals in Eq.~(\ref{eq:tote}) can be obtained analytically~\cite{wp88} as
\begin{equation}
\int^{p_F}_0\!\!\! d \bm p
\int^{p_F}_0\!\!\! d \bm p'
\frac{1}{1+[(\bm p - \bm p')/\mu]^2}
=\frac{32\pi^2}{3}p_F^4\mu^2\biggl[\frac{3}{8}-
\frac{\mu}{2p_F}\arctan \frac{2p_F}{\mu}-
\frac{\mu^2}{16p_F^2}
 + \biggl\{\frac{3}{16}\frac{\mu^2}{p_F^2}+
\frac{1}{64}\frac{\mu^4}{p_F^4}\ln\biggl(
1+\frac{4p_F^2}{\mu^2}\biggr)
\biggr\}
\biggr].
\end{equation}
\end{widetext}
Parameters $\alpha, \beta$ and $ \gamma$ in Eq.~(\ref{eq:tote}) are determined
to reproduce the saturation of the total energy per nucleon
at the normal nuclear density, i.e.
$E/A|_{\rho=\rho_0}= -16\ {\rm MeV}$,
and
$P=\rho^2{\partial (E/A)}/{\partial \rho}|_{\rho=\rho_0} = 0 $ MeV$/$fm$^3$
~\cite{aich2}.
The incompressibility $K$ is obtained from
$K= 9 \rho^2 {\partial^2 (E/A)}/{\partial \rho^2}|_{\rho=\rho_0}$.
Parameters for hard (H) and soft (S) EOS are listed in Table~\ref{parsets}
and the density dependences of the total energy per nucleon
are shown in the right panel of Fig.~\ref{skyeps}.

Parameters $C_\mathrm{ex}^{(k)}$ and $\mu_k$ are taken to reproduce
the real part of the global Dirac optical potential
(Schr\"odinger equivalent potential) of Hama et al.~\cite{Hama1990},
in which
angular distribution and polarization quantities in
 proton--nucleus elastic scatterings
are analyzed in the range of 10 MeV$\sim$1 GeV in Dirac phenomenology.
Single particle potential at $\rho=\rho_0$
\begin{align}
& U(\bm p, \rho_0)=
\alpha +\beta \nonumber\\
&+ \biggl(\frac{4}{3}\pi p_F^3 \biggr)^{-1}
\!\!\!\int^{p_F}_0 \!\!\! d \bm p'
\sum_{k=1,2}
\frac{C_{\rm ex}^{(k)}}{1+[(\bm p - \bm p')/\mu_k]^2}
\nonumber\\
&= \alpha +\beta + \biggl(\frac{4}{3}\pi p_F^3 \biggr)^{-1}
\nonumber\\
&\quad \times \sum_{k=1,2} C_\mathrm{ex}^{(k)}
\pi\mu_k^3\biggl[
\frac{p_F^2+\mu_k^2-p^2}{2p\mu_k}\ln\frac{(p+p_F)^2+\mu_k}{(p-p_F)^2+\mu_k}
 \nonumber\\
&\quad +\frac{2p_F}{\mu_k}
-2 \biggl(
\arctan\frac{p+p_F}{\mu_k}-
\arctan\frac{p-p_F}{\mu_k}
\biggr)
\biggr]
\label{umomint}\ ,
\end{align}
is compared to the Schr\"odinger equivalent potential in Ref.~\cite{Hama1990}
in the left panel of Fig.~\ref{skyeps}.
Parameters for the momentum dependent potentials
are shown as MH and MS in Table~\ref{parsets}.
These parameter sets are based on Ref.~\cite{maru2} 
with simplification
in which Coulomb, surface and Pauli potentials as well as the zero-point
kinetic energy of the Gaussian wave packets are dropped because their
study was focused on nuclear matter below the saturation density.
We have fixed the high energy limit of the optical potential,
$U \to 77$ MeV at $E_\mathrm{inc} \to \infty$,
leading to a constraint $\alpha+\beta=77$ MeV.
This constraint generally makes EOS stiffer
compared to those in Ref.~\cite{Giessen}.

%%%%%%%%%%%%%%%%%%%%%%%%
% RQMD/S
%%%%%%%%%%%%%%%%%%%%%%%%

We include the above MF effects into
{\sc jam}~\cite{jam} by means of simplified RQMD
(RQMD/S)~\cite{RQMD_S} framework.
The Relativistic Quantum Molecular Dynamics ({\sc rqmd})~\cite{sorge,sorge2,maru}
 is a constraint Hamiltonian dynamics,
in which potentials are treated in a covariant way.
RQMD/S~\cite{RQMD_S} uses
much simpler and more practical time fixation constraints
compared to
the original one~\cite{sorge,sorge2,maru}.
For detail, see Appendix A.

In this work,
we take into account potential interactions only between baryons.
Simulation time step size is taken to be $dt=0.1$ fm/$c$
at all incident energies.
We will discuss the influence of
MF for non-nucleonic baryons 
on the flow analysis
and the validity of
this treatment in section~\ref{sec:discussion}.
The violated magnitude of the energy conservation is about 0.4 \%
in average of time and events.

\section{Collective Flows from AGS to SPS Energies}
\label{sec:results}

When two heavy-nuclei collide at high energies at finite impact parameters,
pressure gradient is anisotropic in the initial stages of a collision.
As a result, it generates anisotropic collective flows.
Up to now, several kinds of collective flows are proposed 
to probe high dense matter.
The first one is the sideward flow (also called directed flow)
$\langle p_x\rangle$,
which is defined as the mean value of $p_x$,
where $x$ is defined as the impact parameter direction on the reaction plane.
Sideward flow is mainly generated by the participant-spectator interaction.
Nucleons in the projectile feel repulsive interaction
from the target nucleus during the contact time of projectile and target.
This repulsion pushes projectile nucleons out
in the positive sideward direction if the contact time is long enough.
When the incident energy is very high,
contact time in collisions becomes shorter due to the Lorentz contraction,
therefore sideward flow decreases.
At SPS energies, other types of collective flows,
called as directed ($v_1$) and elliptic ($v_2$) flows,
are mainly measured.
These are defined as the $n$-th Fourier coefficient,
\begin{eqnarray}
{d^3 N \over p_\ssT dp_\ssT dy d\phi}
	&=& {d^2 N \over 2\pi p_\ssT dp_\ssT dy} \nonumber \\
	&& \hspace{-0.5cm} \times \left(
		1 + \sum_n 2 v_n(p_\ssT, y) \cos n\phi
	\right)\ ,
\end{eqnarray}
where the azimuthal angle $\phi$ is measured from the reaction plane.
The directed flow $v_1$ is the first Fourier coefficient
of the azimuthal distribution
\begin{equation}
v_1 =\langle\cos\phi\rangle = \biggl\langle{p_x \over p_\ssT}\biggr\rangle\ ,
\end{equation}
and the elliptic flow $v_2$
are the second Fourier coefficient of the
azimuthal distribution
\begin{equation}
v_2 =\langle\cos 2\phi\rangle
	= \biggl\langle{p_x^2 - p_y^2 \over p_\ssT^2 }\biggr\rangle.
\end{equation}
These collective flows are reviewed in Ref.~\cite{sissps}.

The effects of MF in high-energy heavy-ion collisions
are visible but not very large in single particle spectra,
such as rapidity distribution $dN/dy$
or transverse mass distribution $d^2N/m_\ssT dm_\ssT dy$.
In this section, we demonstrate that 
MF effects are essential to study anisotropic collective flows
in the hadron-string transport model {\sc jam} with MF potentials.

\subsection{Collective Flows at AGS Energies}

%%%%%%%%%%%%%%%%%%%%%%%%%%%%%%%%%%%%%%%%%%%%%%%%%%%%%%%%%%%%%%%%%%%%%%
%  2, 4, 6, 8 GeV rapidity dependence of the sideward flow
%%%%%%%%%%%%%%%%%%%%%%%%%%%%%%%%%%%%%%%%%%%%%%%%%%%%%%%%%%%%%%%%%%%%%%
\begin{figure*}[tbh]
\includegraphics[width=10cm,clip]{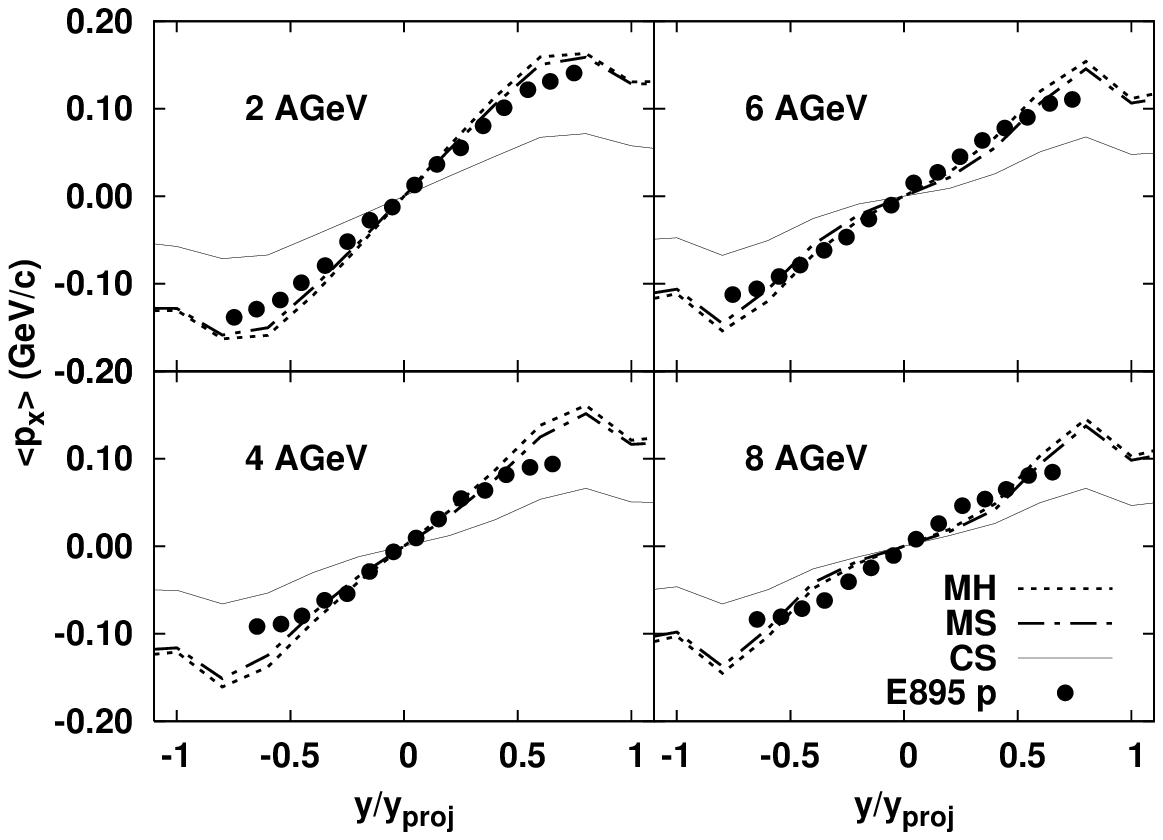}~\hspace{-2cm}
\includegraphics[width=10cm,clip]{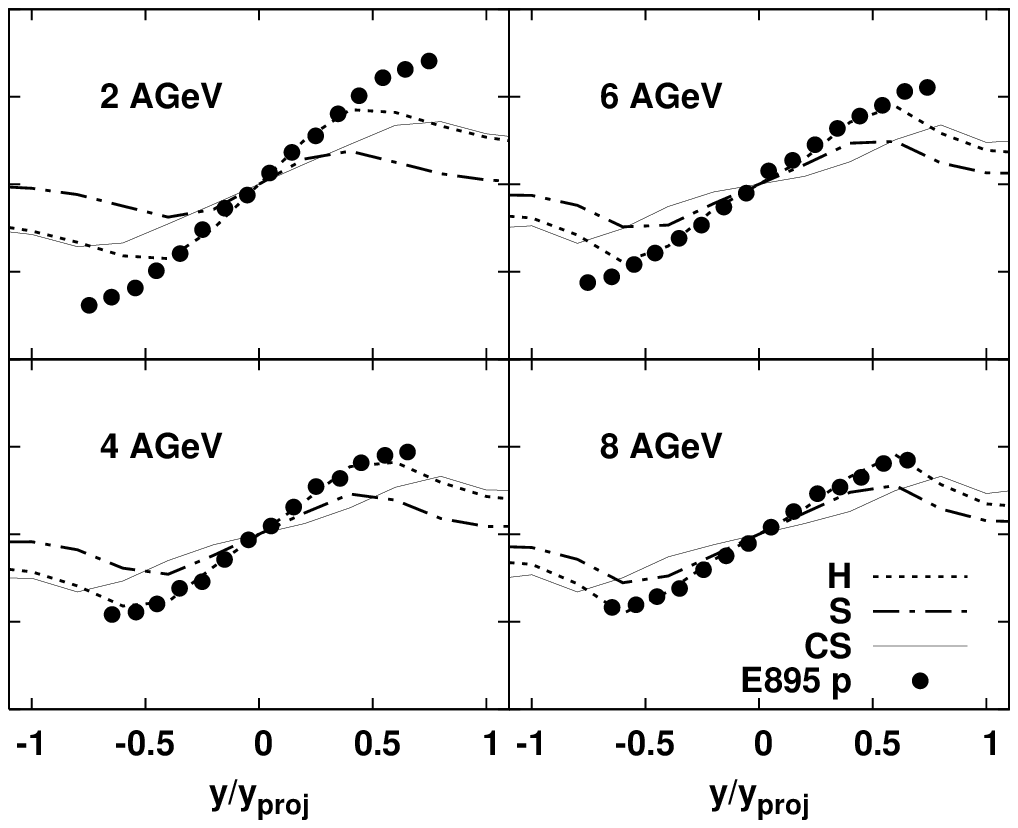}
\caption{
Sideward flows $\langle p_x \rangle$
of protons in mid-central Au+Au collisions at $(2-8)A$ GeV
are compared to the AGS-E895 data~\cite{E895-00}.
Lines show
the calculated results of
Cascade with momentum dependent hard/soft mean-field (MH/MS, left panels),
Cascade with momentum independent mean-field (H/S, right panels)
and Cascade without mean-field (CS).
The experimental data are shown in both of the left and right panels.
}
\label{ypx2-8}
\end{figure*} 

%%%%%%%%%%%%%%%%%%%%%%%%%%%%%%%%%%%%%%%%%%%%%%%%%%%%%%%%%%%%%%%%%%%%%%
%  11 GeV rapidity dependence of the sideward flow
%%%%%%%%%%%%%%%%%%%%%%%%%%%%%%%%%%%%%%%%%%%%%%%%%%%%%%%%%%%%%%%%%%%%%%
\begin{figure*}[tbh]
\begin{center}
\includegraphics[width=9cm,clip]{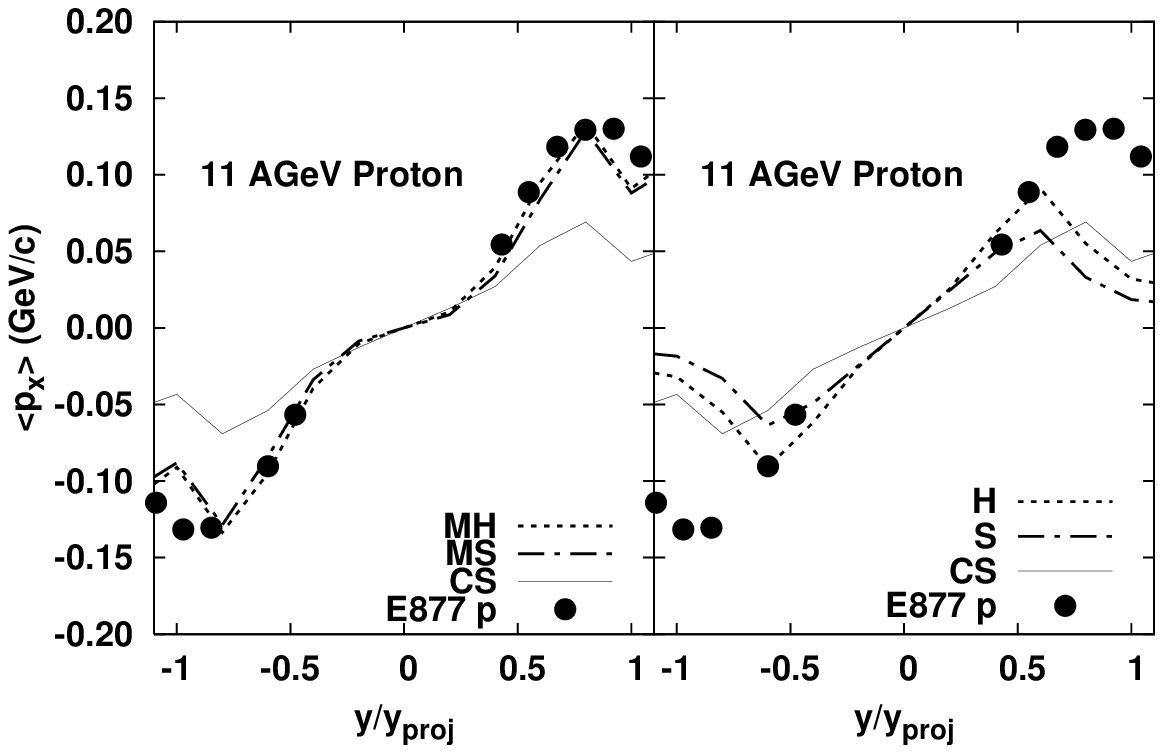}~\hspace{-1cm}
\includegraphics[width=9cm,clip]{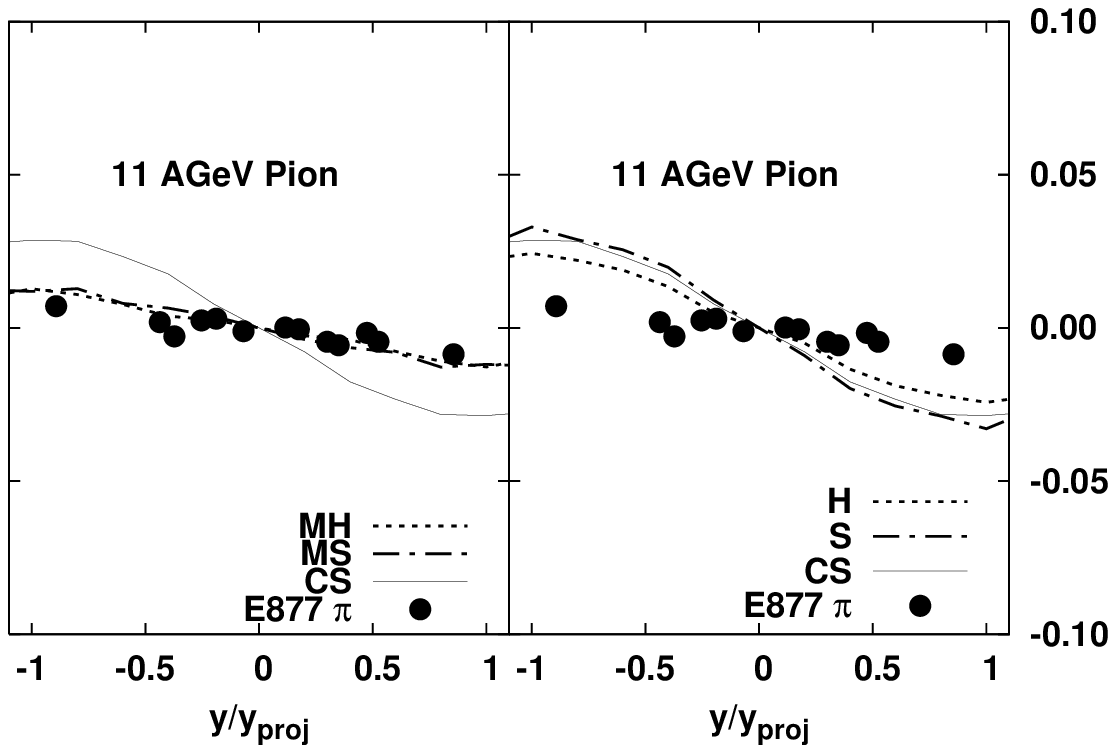}
\caption{
Comparison of calculated sideward flow $\langle p_x \rangle$
of protons (left) and pions (right) in $11A$ GeV Au+Au mid-central collisions
to AGS-E877 data~\cite{E877}.
The meaning of the lines is the same as Fig.~\ref{ypx2-8}.
}\label{ypx11}
\end{center}
\end{figure*} 
%%%%%%%%%%%%%%%%%%%%%%%%%%%%%%%%%%%%%%%%%%%%%%%%%%%%%%%%%%%%%%%%%%%%%%%%%%%

%%%%%%%%%%%%%%%%%%%%%%%%%%%%%%%%%%%%%%%%%%%%%%%%%%%%%%%%%%%%%%%%%%%%%%%%%
%  <v2(pt)>  at 2, 4, 6GeV
%%%%%%%%%%%%%%%%%%%%%%%%%%%%%%%%%%%%%%%%%%%%%%%%%%%%%%%%%%%%%%%%%%%%%%%%%
\begin{figure*}[tbh]
\hspace*{-0.7cm}
\includegraphics[width=10cm,clip]{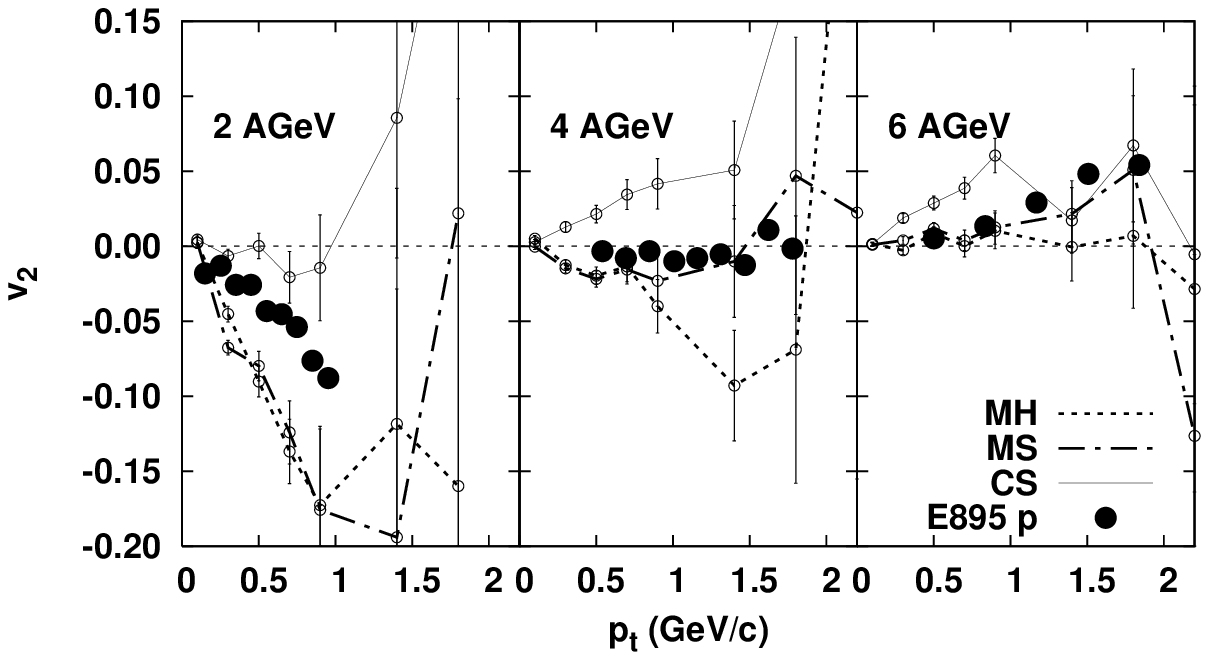}~\hspace{-1cm}
\includegraphics[width=10cm,clip]{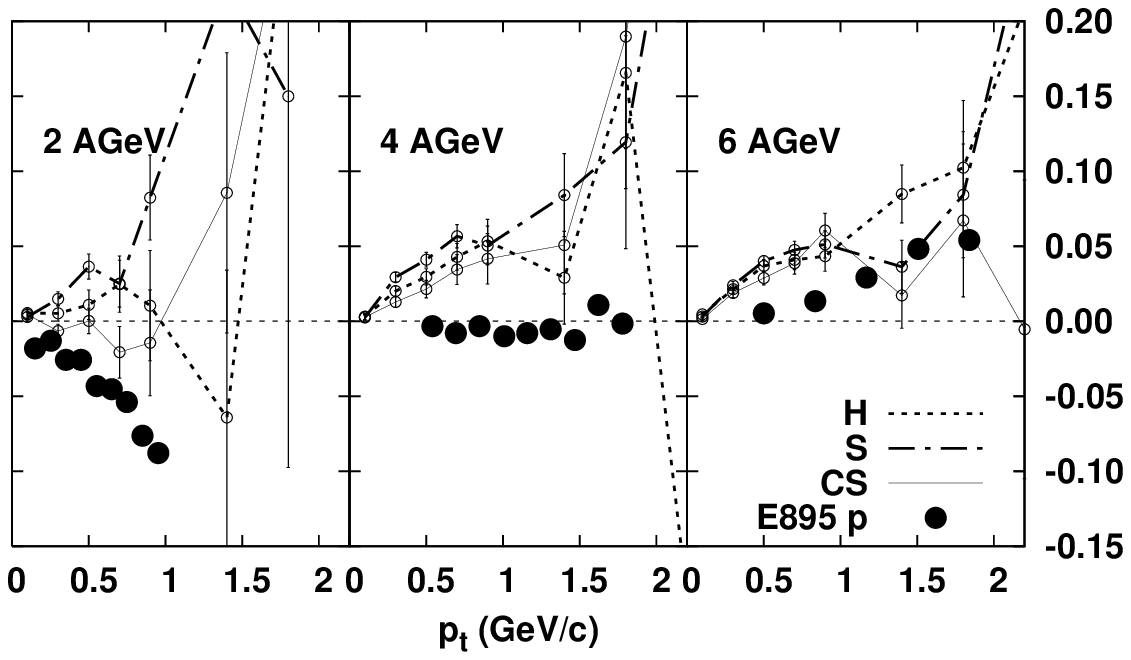}
\vspace{-2cm}
\caption{Transverse momentum dependence of the
elliptic flow $v_2$
for protons
in Au+Au mid-central collisions
at $(2,4,6)A$ GeV
are compared to AGS-E895 data~\cite{E895-00}.
The meaning of the lines is the same as Fig.~\ref{ypx2-8}.
}\label{agsptv2}
\end{figure*} 
%%%%%%%%%%%%%%%%%%%%%%%%%%%%%%%%%%%%%%%%%%%%%%%%%%%%%%%%%%%%%%%%%%%%%%%%%

We show proton sideward flow $\langle p_x\rangle$
in mid-central Au+Au collisions
at AGS energies ($E_\mathrm{inc} = (2-11)A$ GeV)
together with AGS-E895 data~\cite{E895-00}
in Fig.~\ref{ypx2-8} and in the left panel of Fig.~\ref{ypx11}.
We choose the impact parameter range $4<b<8$ fm
in the calculations
which roughly corresponds
to mid-central collisions in experimental data.

It is seen that
both Cascade and momentum independent soft (S) MF results
are inconsistent with the data.
The magnitude of $\langle p_x\rangle$ in forward rapidity region
($y/y_{\rm proj}\simeq \pm 1$) is small compared to the data,
and the slope parameters at mid-rapidity are also smaller
than that of the data
with soft MF.
The momentum independent soft MF reduces $\langle p_x\rangle$
in forward rapidity region,
and enhances the slope parameters at mid-rapidity.
The former is an unfavorable effect in explaining the data,
and the latter is not enough.
With momentum independent hard (H) MF,
slope parameter is well reproduced,
but the $\langle p_x\rangle$ at forward rapidities
are smaller than the data
especially at $E_\mathrm{inc} = 2$ and $11 A$GeV.

Proton sideward flow data are 
qualitatively
reproduced with the momentum dependent MF.
The momentum dependent MF pushes up the flow 
almost
linearly as a function of rapidity,
and it becomes 
closer to 
the data,
while the $\langle p_x \rangle$ values at forward rapidities
may be a little too large compare to the data
at $E_\mathrm{inc}=4-8A$ GeV.
As the incident energy increases,
MF effects on the slope parameter at mid-rapidity become small,
but we can still see clear differences at forward rapidities
between the results with and without momentum dependence.

Our results suggest the necessity of
the momentum dependence in the MF 
to give large magnitude emission to $x$ direction at forward rapidity.
We note that our results with momentum dependent MF are consistent 
with the previous calculations with  MF 
on the collective flow data
at AGS energies~\cite{sc02,sahu2,da02}
as well as SIS energies~\cite{persram,fopid04}.

The importance of the momentum dependence in the MF
is 
also
seen in the transverse momentum dependence
of the proton $v_2$ as shown in Fig.~\ref{agsptv2}.
Only if momentum dependence is included, 
we reproduce the
strong squeezing
at $E_\mathrm{inc}=2A$ GeV.
of the $p_\ssT$ dependence.

In the right panel of
Fig.~\ref{ypx11}, we plot the results of sideward flow $\langle p_x\rangle$
for pions in Au+Au collisions at $E_\mathrm{inc} = 11A$ GeV.
The sideward flow $\langle p_x\rangle$ of 
pions are suppressed 
significantly
by
momentum dependent
 MF.
This may be because pions are
trailed by nucleons which is affected by MF,
giving visible differences.

\subsection{Directed Flow at SPS energies}

%%%%%%%%%%%%%%%%%%%%%%%%%%%%%%%%%%%%%%%%%%%%%%%%%%%%%%%%%%%%%%%%%%%%%%%%%%%%
%   v_1(y) at 40 and 158 GeV
%%%%%%%%%%%%%%%%%%%%%%%%%%%%%%%%%%%%%%%%%%%%%%%%%%%%%%%%%%%%%%%%%%%%%%%%%%%%
\begin{figure*}[thb]
\includegraphics[width=9.5cm]{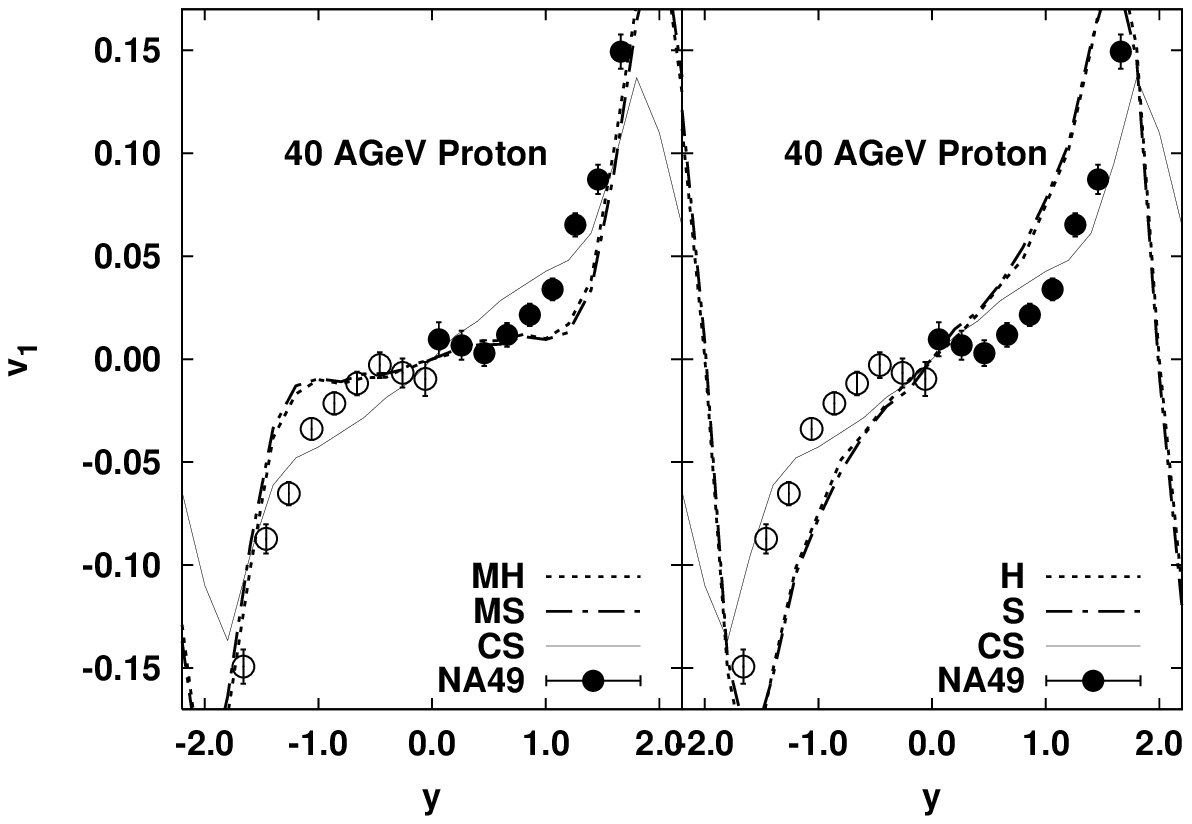}~\hspace{-1cm}
\includegraphics[width=9.5cm]{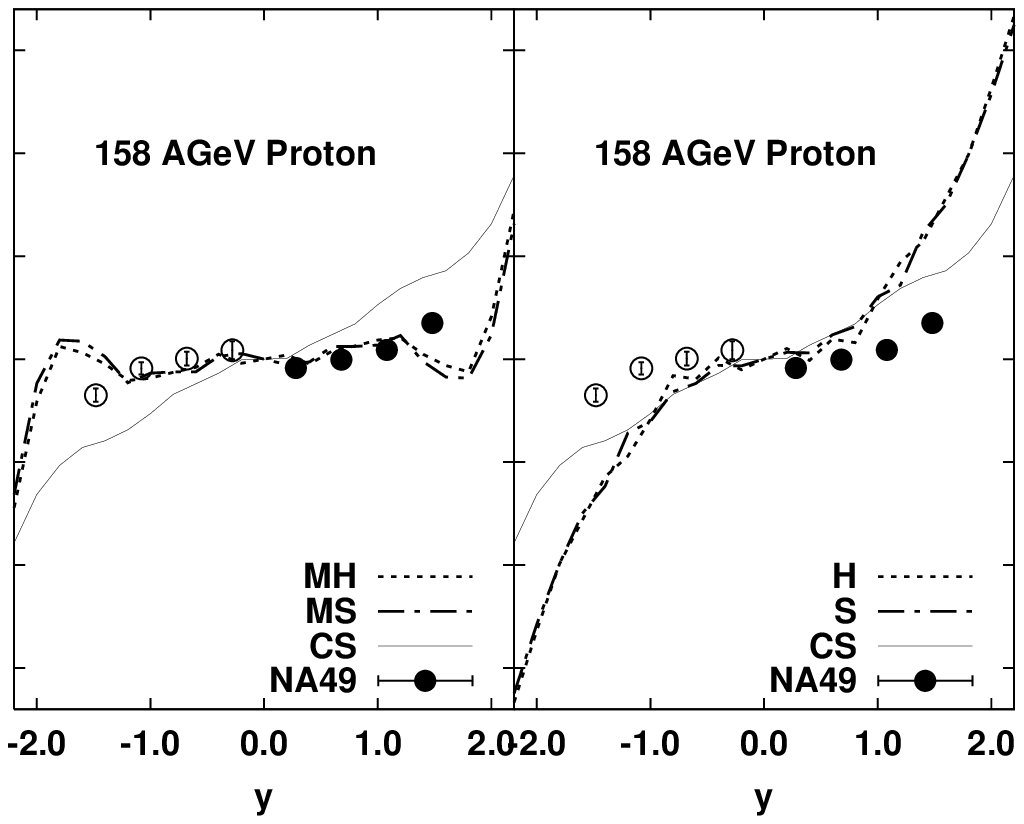}
\caption{
Proton directed flows $v_1$ as a function of rapidity
in mid-central Pb+Pb collisions
at $E_\mathrm{inc}=40A$ GeV (left panel) and $158A$ GeV (right panel)
in comparison with SPS-NA49 data~\cite{NA49-03}.
Lines show
the calculated results of
Cascade with momentum dependent hard/soft mean-field (MH/MS),
Cascade with momentum independent mean-field (H/S)
and Cascade without mean-field (CS).
}
\label{yv1p}
\end{figure*} 

%%%%%%%%%%%%%%%%%%%%%%%%%%%%%%%%%%%%%%%%%%%%%%%%%%%%%%%%%%%%%%%%%%%%%%%%%%%%
%   v_1(pt) for p  at 40 and 158AGeV
%%%%%%%%%%%%%%%%%%%%%%%%%%%%%%%%%%%%%%%%%%%%%%%%%%%%%%%%%%%%%%%%%%%%%%%%%%%%
\begin{figure*}[thb]
\includegraphics[width=9.5cm]{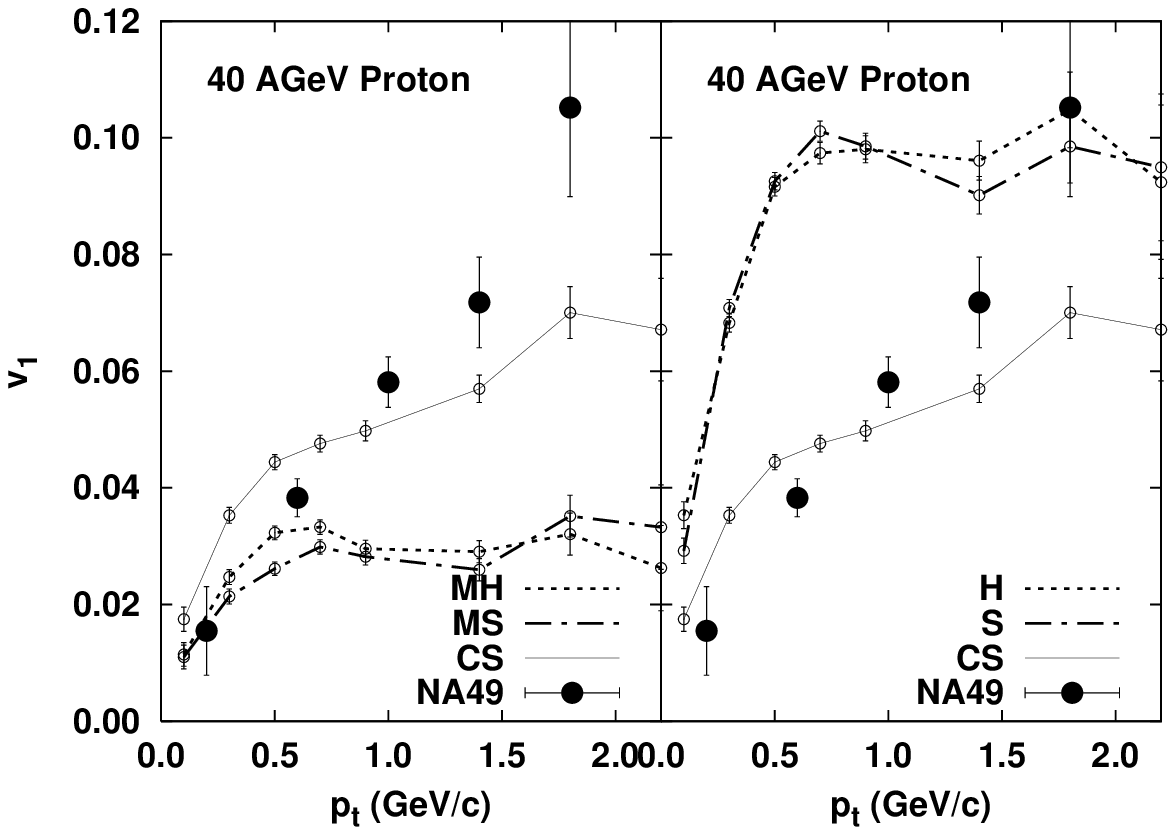}~\hspace{-1cm}
\includegraphics[width=9.5cm]{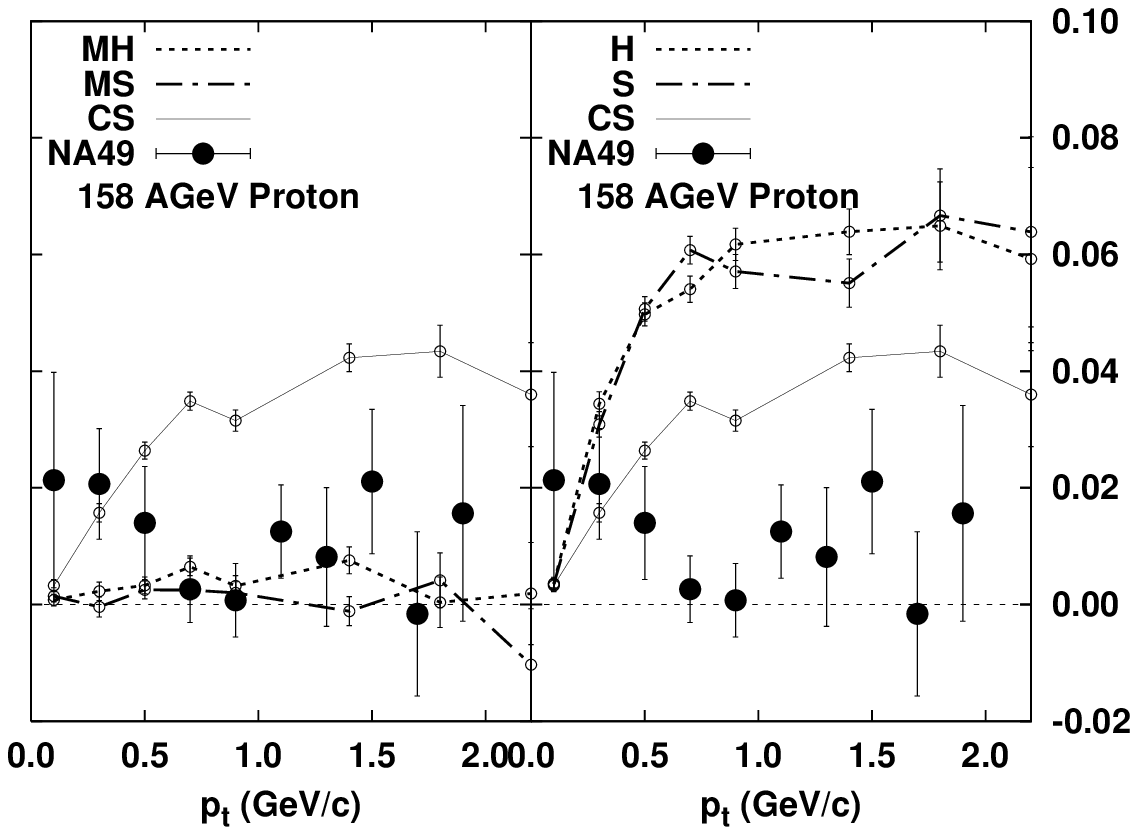}
\caption{
Proton directed flows $v_1$ as a function of transverse momentum
in mid-central Pb+Pb collisions
at $E_\mathrm{inc}=40A$ GeV (left panel) and $158A$ GeV (right panel)
are compared with SPS-NA49 data~\cite{NA49-03}.
The meaning of the lines is the same as Fig.~\ref{yv1p}.
}
\label{ptv1p}
\end{figure*} 

%%%%%%%%%%%%%%%%%%%%%%%%%%%%%%%%%%%%%%%%%%%%%%%%%%%%%%%%%%%%%%%%%%%%%%%%%%%%
%   v_1(y) for pi  at 40 and 158AGeV
%%%%%%%%%%%%%%%%%%%%%%%%%%%%%%%%%%%%%%%%%%%%%%%%%%%%%%%%%%%%%%%%%%%%%%%%%%%%
\begin{figure*}[thb]
\includegraphics[width=9.5cm]{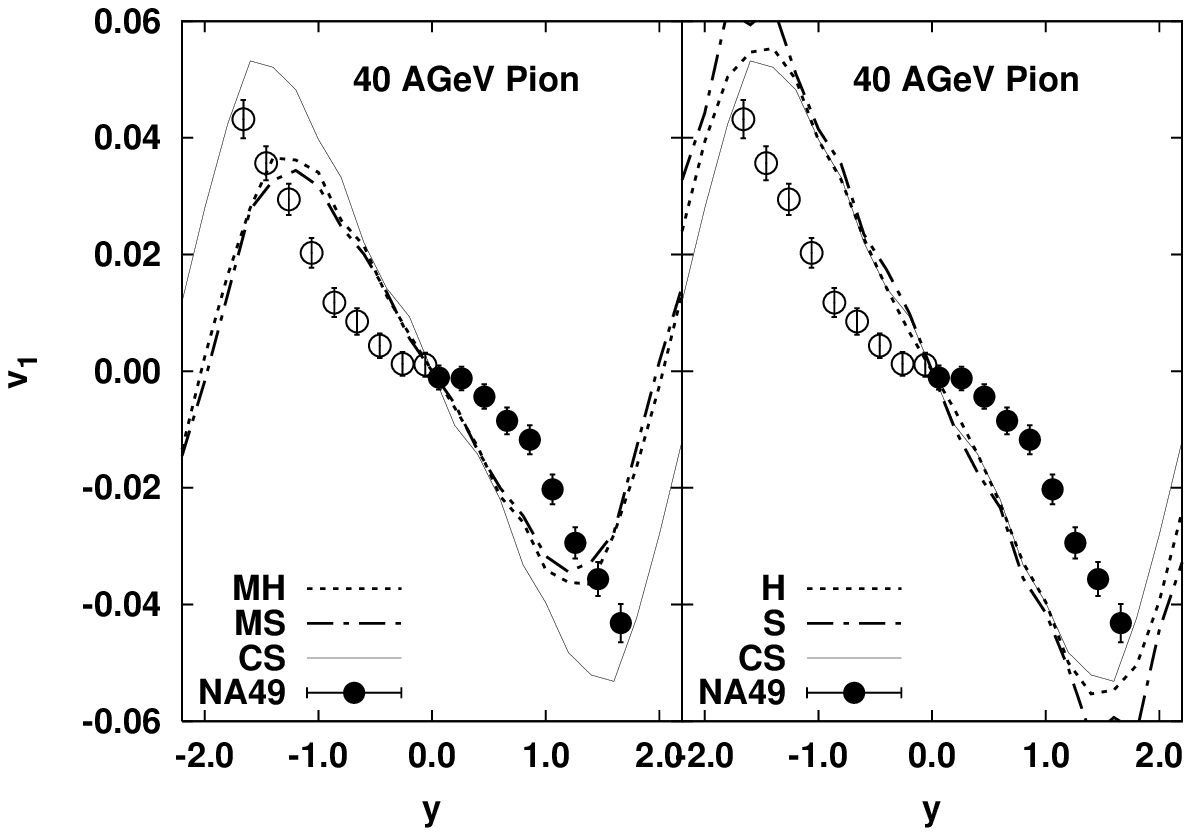}~\hspace{-1cm}
\includegraphics[width=9.5cm]{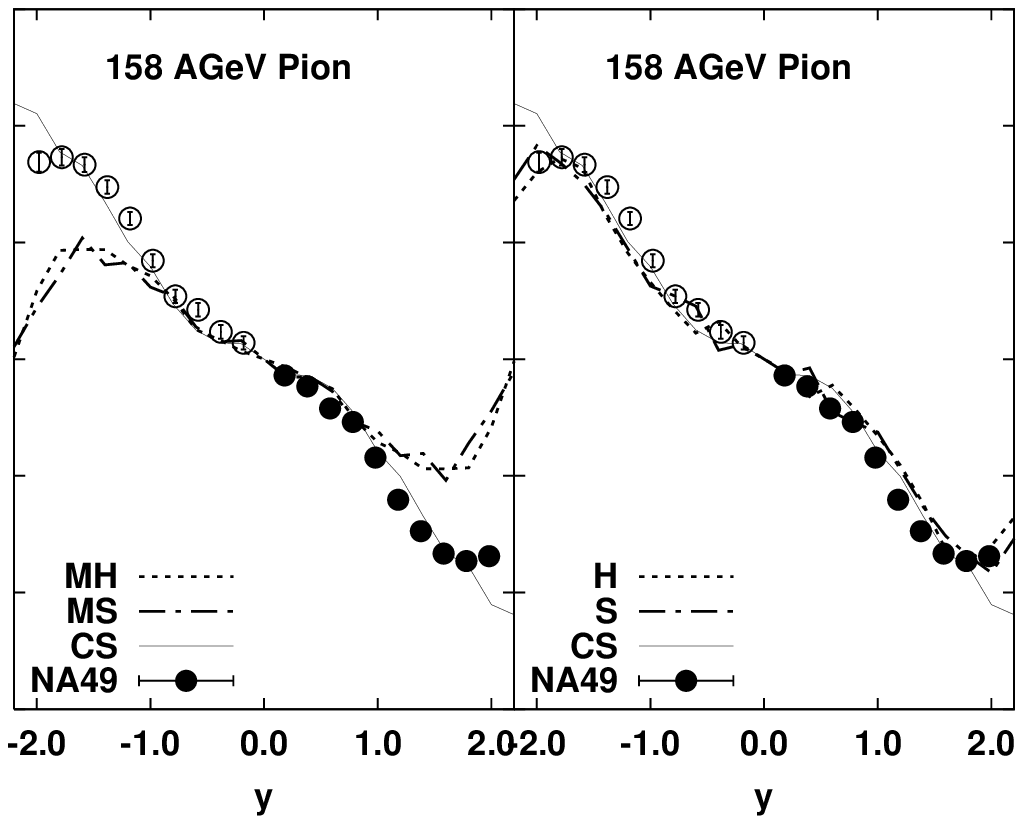}
\caption{
Pion directed flows $v_1$ as a function of rapidity
in mid-central Pb+Pb collisions
at $E_\mathrm{inc}=40A$ GeV (left panel) and $158A$ GeV (right panel)
are compared with SPS-NA49 data~\cite{NA49-03}.
The meaning of the lines is the same as Fig.~\ref{yv1p}.
}
\label{yv1pi}
\end{figure*}

%%%%%%%%%%%%%%%%%%%%%%%%%%%%%%%%%%%%%%%%%%%%%%%%%%%%%%%%%%%%%%%%%%%%%%%%%%%%
%   v_1(pt) for pi  at 40 and 158AGeV
%%%%%%%%%%%%%%%%%%%%%%%%%%%%%%%%%%%%%%%%%%%%%%%%%%%%%%%%%%%%%%%%%%%%%%%%%%%%
\begin{figure*}[thb]
\includegraphics[width=9.5cm]{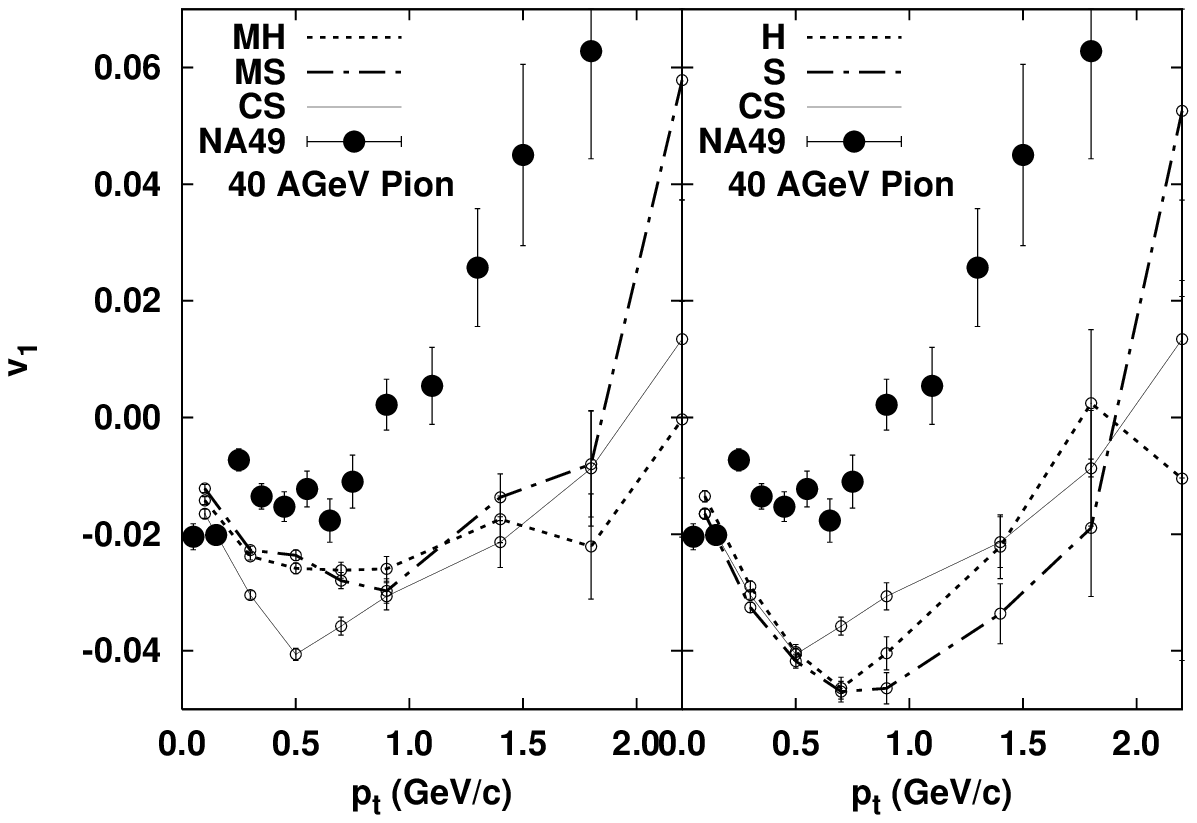}~\hspace{-1cm}
\includegraphics[width=9.5cm]{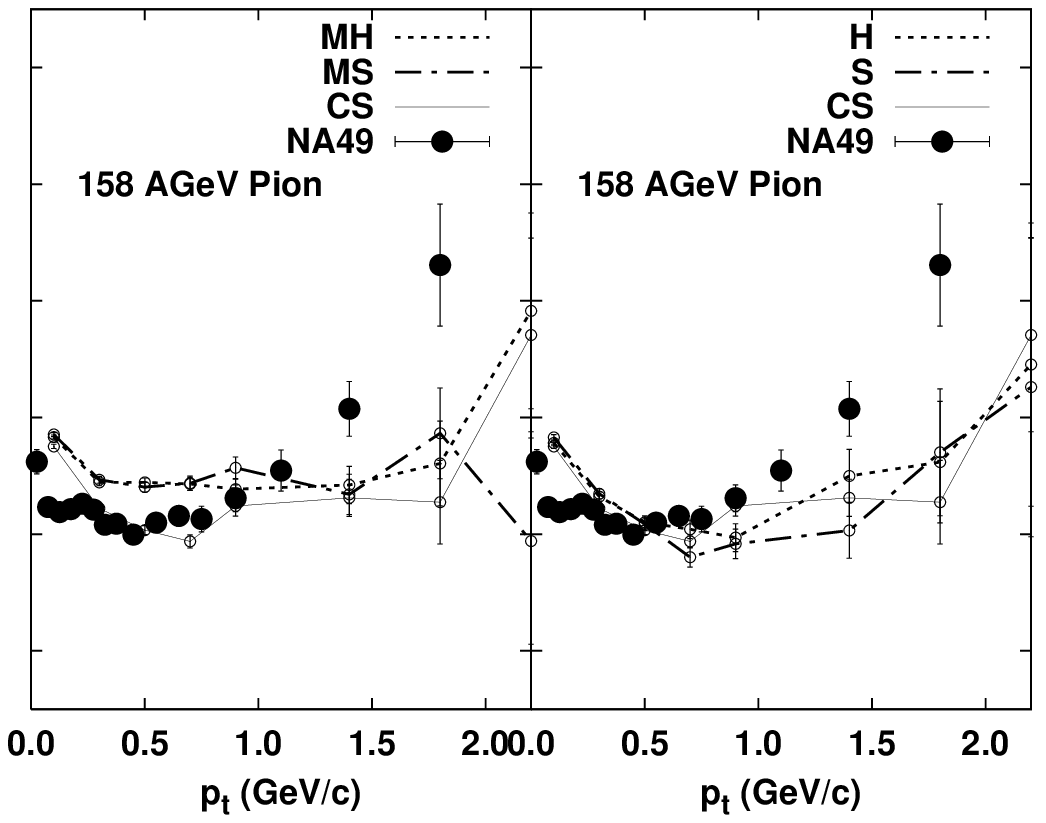}
\caption{
Pion directed flows $v_1$ as a function of transverse momentum
for $0<y<1.5$ in Pb+Pb collisions
at $E_\mathrm{inc}= 40A$ GeV (left panel) and $158A$ GeV (right panel)
are compared with SPS-NA49 data~\cite{NA49-03}.
The meaning of the lines is the same as Fig.~\ref{yv1p}.
}
\label{ptv1pi}
\end{figure*}

Directed flow $v_1$ has been measured
at SPS energies ($E_{\rm inc} = (40,158)A$ GeV)
instead of $\langle{p_x}\rangle$ as a function of rapidity.
In Fig.~\ref{yv1p},
we compare the rapidity dependence of proton $v_1$
with the data
in mid-central Pb+Pb collisions at $E_{\rm inc} =40A$ and $158A$ GeV
by CERN-NA49 collaboration~\cite{NA49-03},
both of which are deduced by the reaction plane method (standard method).
One can see that
momentum dependent MF generally improves the description of $v_1$.

%
% Protons
%
It is interesting to note that the Cascade model overestimates $v_1$ 
for protons
contrary to the underestimate of $\langle p_x\rangle$ at AGS energies.
We also see that $v_1$ is reduced at SPS energies with momentum dependent MF,
while $\langle p_x \rangle$ is enhanced at AGS energies.
This is a reverse behavior compared to that at lower incident energies.
Note also that the results with momentum independent MF 
predict larger $v_1$ than that of the Cascade results.

In Fig.~\ref{yv1p},
the results from the momentum dependent MF show a
flat behavior at mid-rapidity at $158A$ GeV.
The `wiggle'
(a negative slope of the proton $v_1$ near mid-rapidity)~\cite{WIG}
has been reported at peripheral collisions~\cite{NA49-03}.
It is interesting to study this in detail in the future.

In Fig.~\ref{ptv1p},
we compare transverse momentum dependence of $v_1$ for protons
in Pb+Pb collisions at $E_\mathrm{inc}=40A$ and $158A$ GeV with the data.
We choose rapidity cut
$|y|<1.8$ for 40$A$ GeV and $|y|<2.1$ for 158$A$ GeV
according to the experimental cuts.
The $p_\ssT$ dependence at 158$A$ GeV is very different from that at 40$A$ GeV. 
Dense baryonic matter is tentatively formed in the calculations
up to around 40$A$ GeV,
while many strings are formed and hadrons are
formed later at 158$A$ GeV at mid-rapidity.
As a result, $v_1$ does not necessarily grow
as a function of $p_\ssT$ at 158$A$ GeV,
because strings do not feel MF in our model,
and hadrons with large $p_\ssT$ from string decay
have long formation time in the total CM system,
and they would have smaller chances to interact
with other hadrons before strings decay.

%
% Pions
%
Let us now turn to the pion $v_1$.
we show $v_1$ for pions as a function of
rapidity in Fig.~\ref{yv1pi}
and transverse momentum in Fig.~\ref{ptv1pi} at $40A$ and $158A$ GeV.
It is seen that MF effects for pion $v_1$ are very small
especially at mid-rapidities.
MF effects are only seen in the forward rapidity region for momentum
dependent MF.
At $40A$ GeV, in the forward rapidity region, we find
reduction (enhancement) of $v_1$ in
the momentum dependent (independent) MF 
results compared to Cascade ones.
This comes from the counteraction from protons;
momentum dependent (independent) MF
reduces (enhances) proton $v_1$ in the mid-rapidity region,
and pion $v_1$ anti-correlates with proton $v_1$.
Probably, we need to include pion MF for a better understanding
of the collective flows at SPS energies.

\subsection{Elliptic Flow at SPS Energies}

Since $v_1$ signal becomes small due to the short participant-spectator
interaction time at high energies,
the next Fourier coefficient, called as the elliptic flow $v_2$,
has been discussed more extensively at SPS and RHIC.
At these energies,
the participants form an almond-like shape in the transverse plane
after the spectators go through,
and this almond shaped participants start to expand more strongly 
in the $x$ (shorter axis of the almond) direction,
due to the higher pressure gradient
if the participants are well thermalized.
This expansion is known to lead to the enhancement
of in-plane particle emission,
i.e. positive elliptic flow $v_2$.

\begin{figure*}[thb]
\includegraphics[width=9cm]{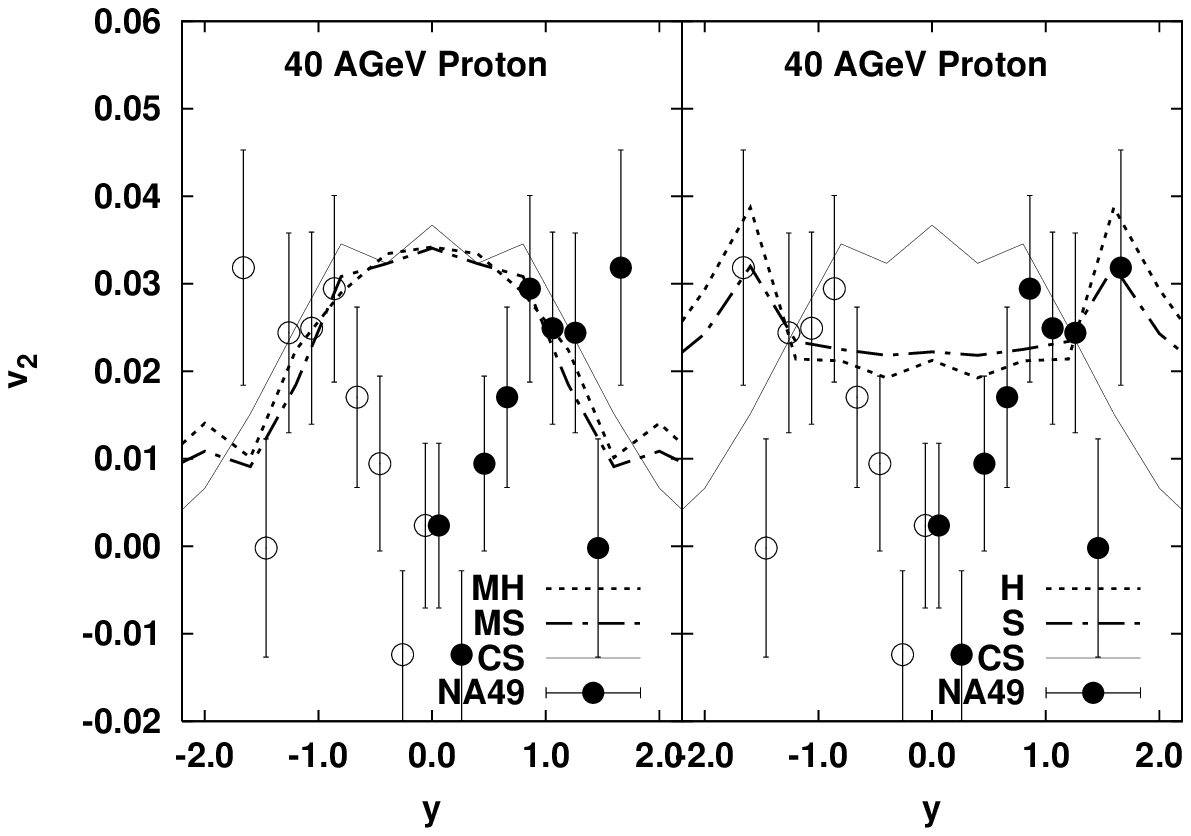}~\hspace{-1cm}
\includegraphics[width=9cm]{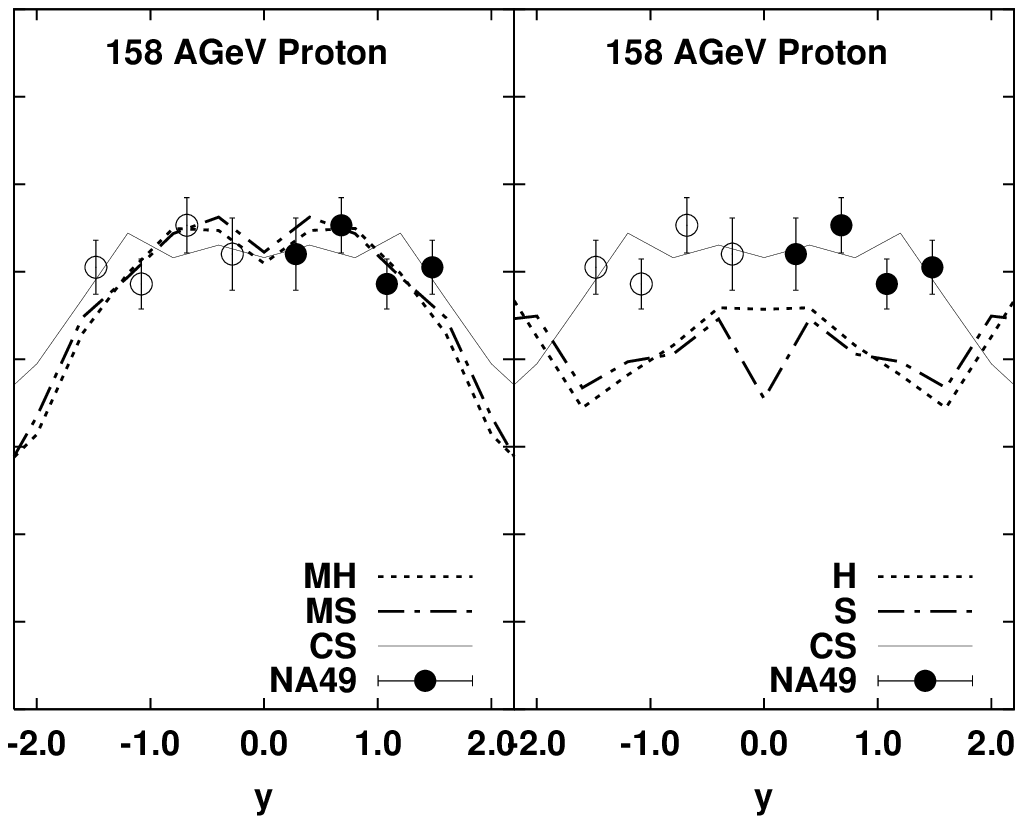}
\caption{
Proton elliptic flows $v_2$ as a function of rapidity
in mid-central Pb+Pb collisions
at $40A$ GeV (left) and $158A$ GeV (right).
Lines show
the calculated results of
Cascade with momentum dependent hard/soft mean-field (MH/MS),
Cascade with momentum independent mean-field (H/S)
and Cascade without mean-field (CS).
Experimental data are taken from SPS-NA49~\cite{NA49-03}.
}
\label{yv2p}
\end{figure*} 
\begin{figure*}[thb]
\includegraphics[width=9cm]{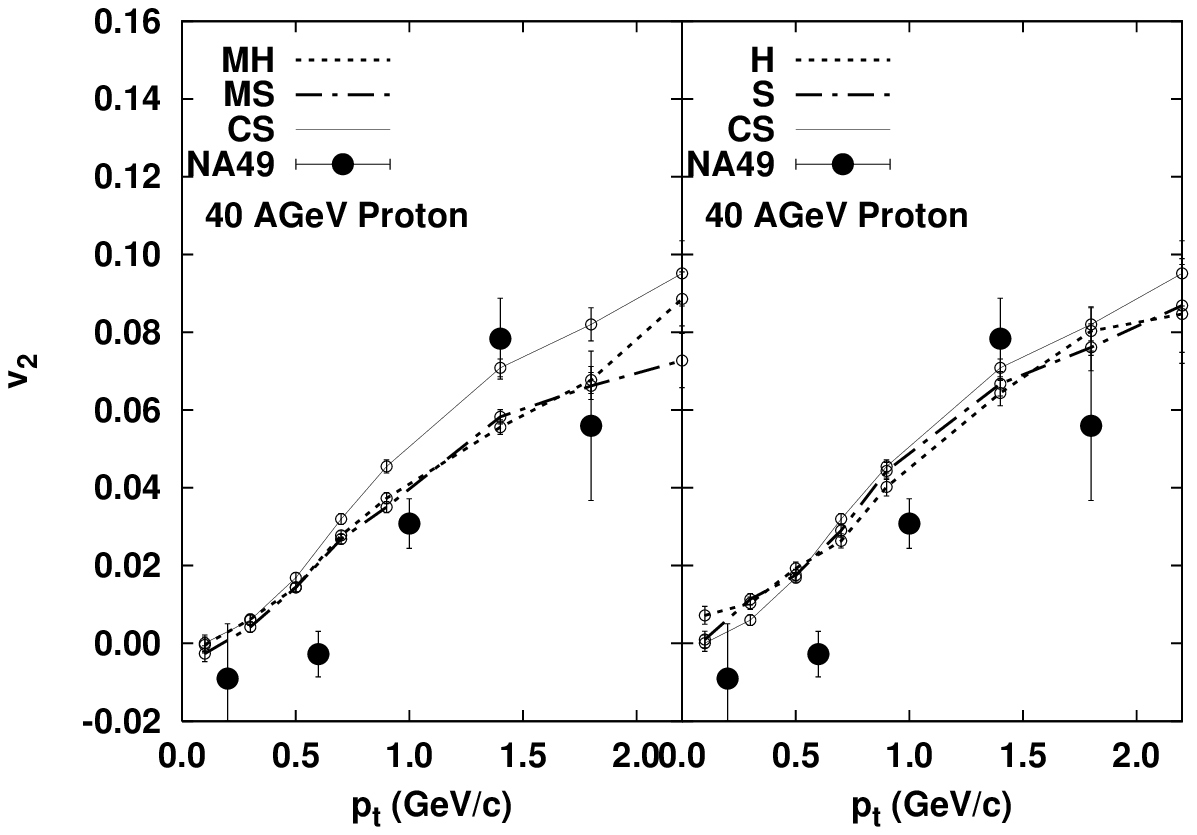}~\hspace{-1cm}
\includegraphics[width=9cm]{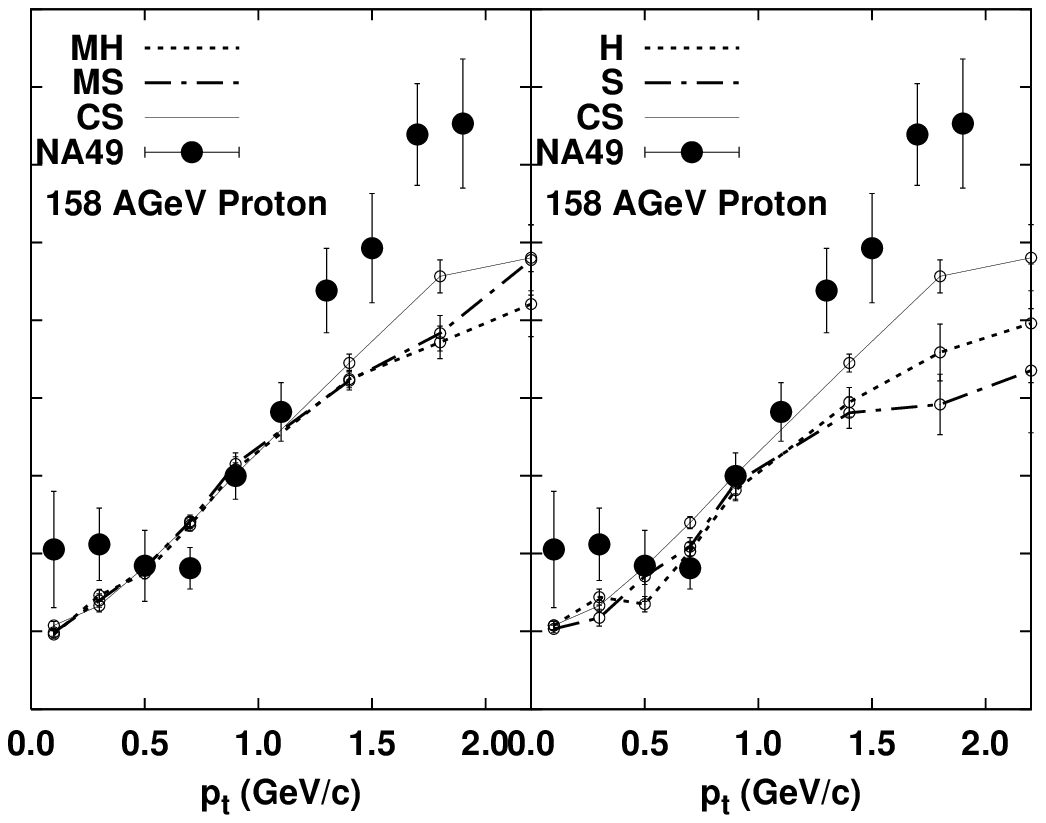}
\caption{
Proton Elliptic flows $v_2$ as a function of transverse momentum.
The meaning of the lines is the same as Fig.~\ref{yv2p}.
}
\label{ptv2p}
\end{figure*} 

In Figs.~\ref{yv2p} and \ref{ptv2p}, 
we plot
the results of
the rapidity and transverse momentum dependence of $v_2$
for protons
at SPS energies (40$A$ and 158$A$ GeV)
together with the SPS-NA49 data~\cite{NA49-03}.
At SPS energies,
the Cascade model generally explains the proton $v_2$ data qualitatively,
including
the flat behavior of $v_2(y)$ at mid-rapidities at 158$A$ GeV
and the approximate linear $p_\ssT$ dependence of $v_2(p_\ssT)$.
One exception is the missing collapse of $v_2(y)$ at mid-rapidity
at 40$A$ GeV.
This collapse seen in the NA49 data may be an 
indication of a first order phase transition
at high baryon densities achieved
in the Pb+Pb collisions at 40$A$ GeV~\cite{stoecker2004}.

Effects of MF
are small for proton $v_2$ at SPS energies.
Elliptic flow is most easily generated in the
early stages of the collisions, since spatial anisotropy
is the largest.
However, at SPS energies, 
string excitations dominate particle production
at early times in the model and
those strings are not affected by the nuclear mean field.
That is the reason why MF effects are small at SPS energies
in our results.

\begin{figure*}[thb]
\includegraphics[width=9cm]{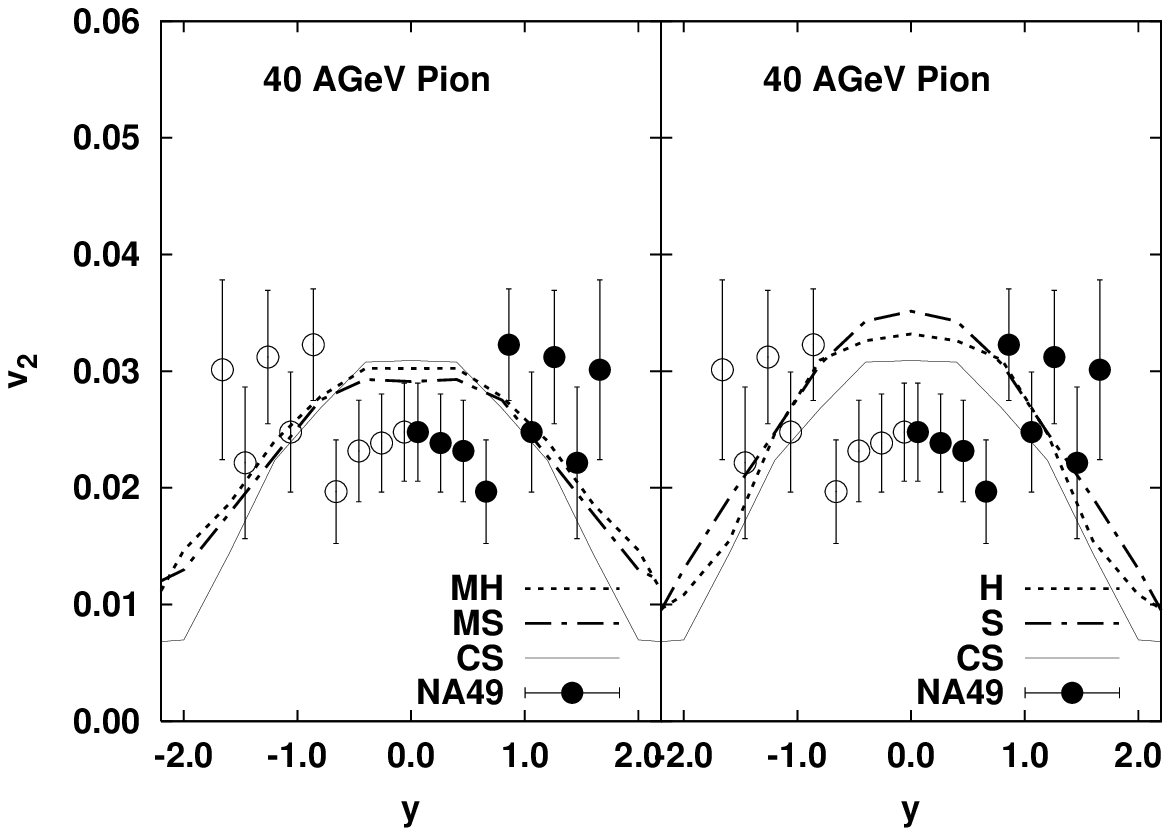}~\hspace{-1cm}
\includegraphics[width=9cm]{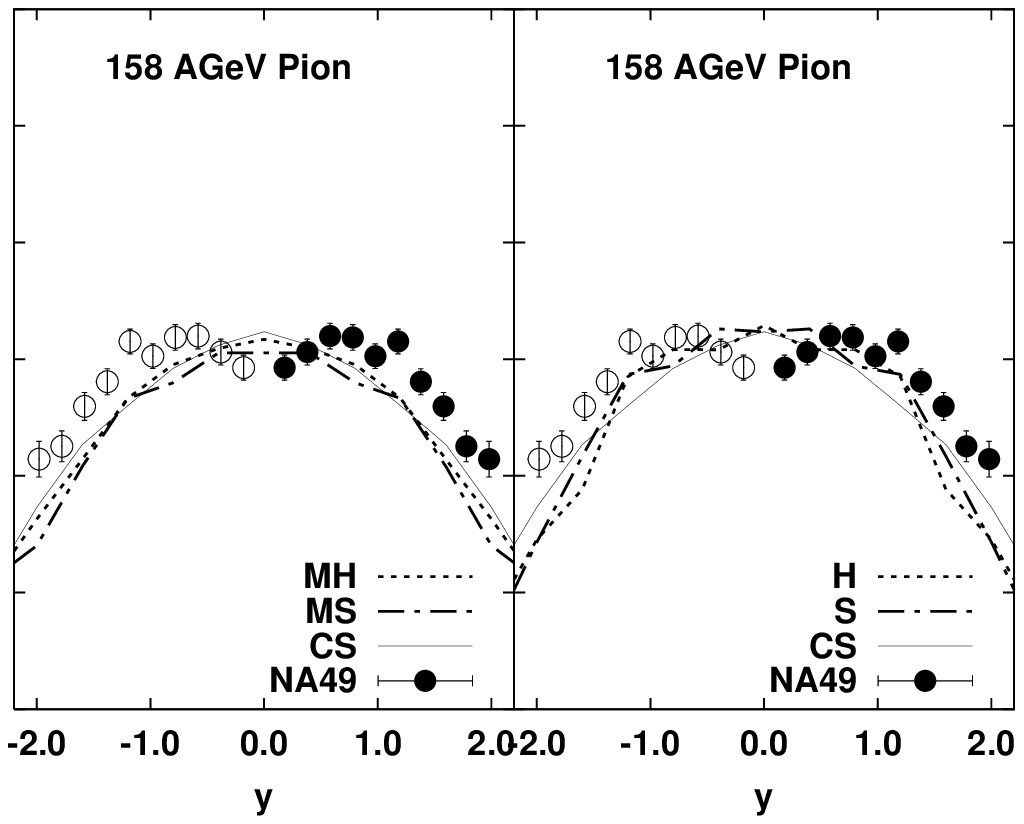}
\caption{
Pion elliptic flows $v_2$ as a function of rapidity.
The meaning of the lines is the same as Fig.~\ref{yv2p}.
}
\label{yv2pi}
\end{figure*}

\begin{figure*}[thb]
\includegraphics[width=9cm]{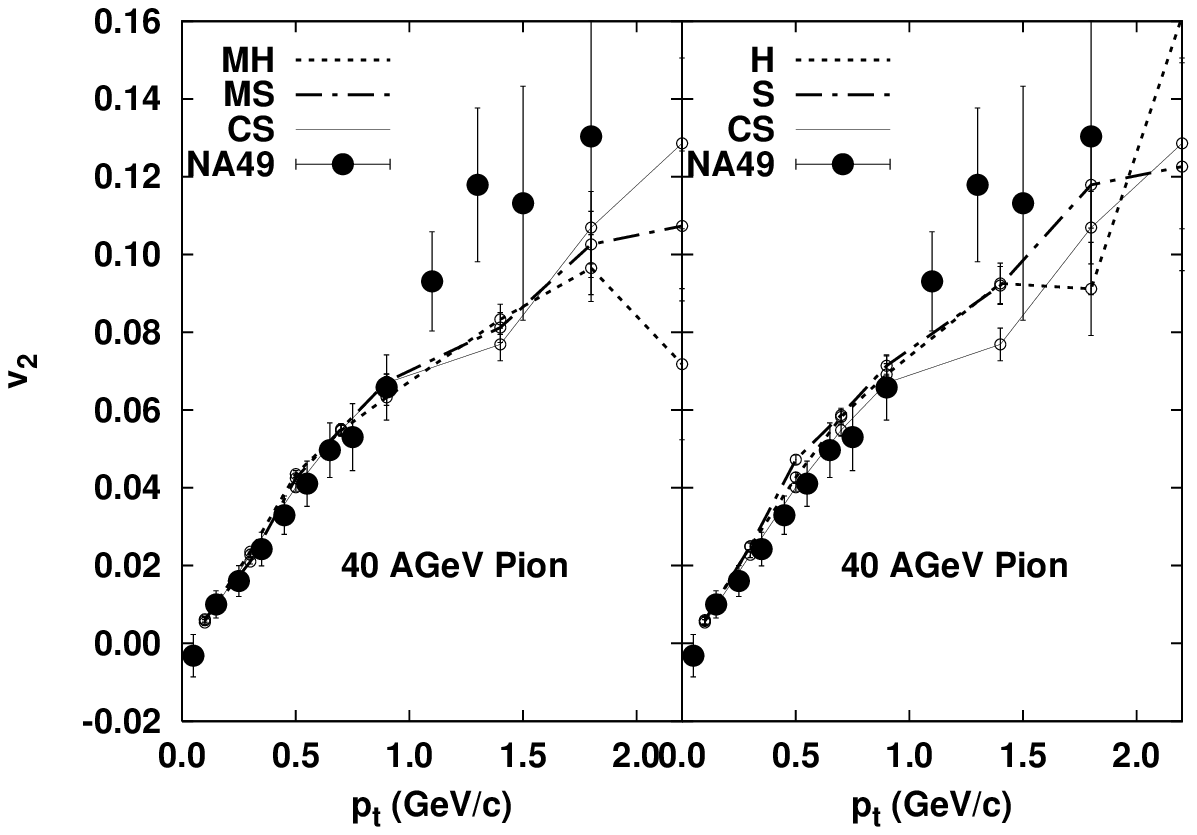}~\hspace{-1cm}
\includegraphics[width=9cm]{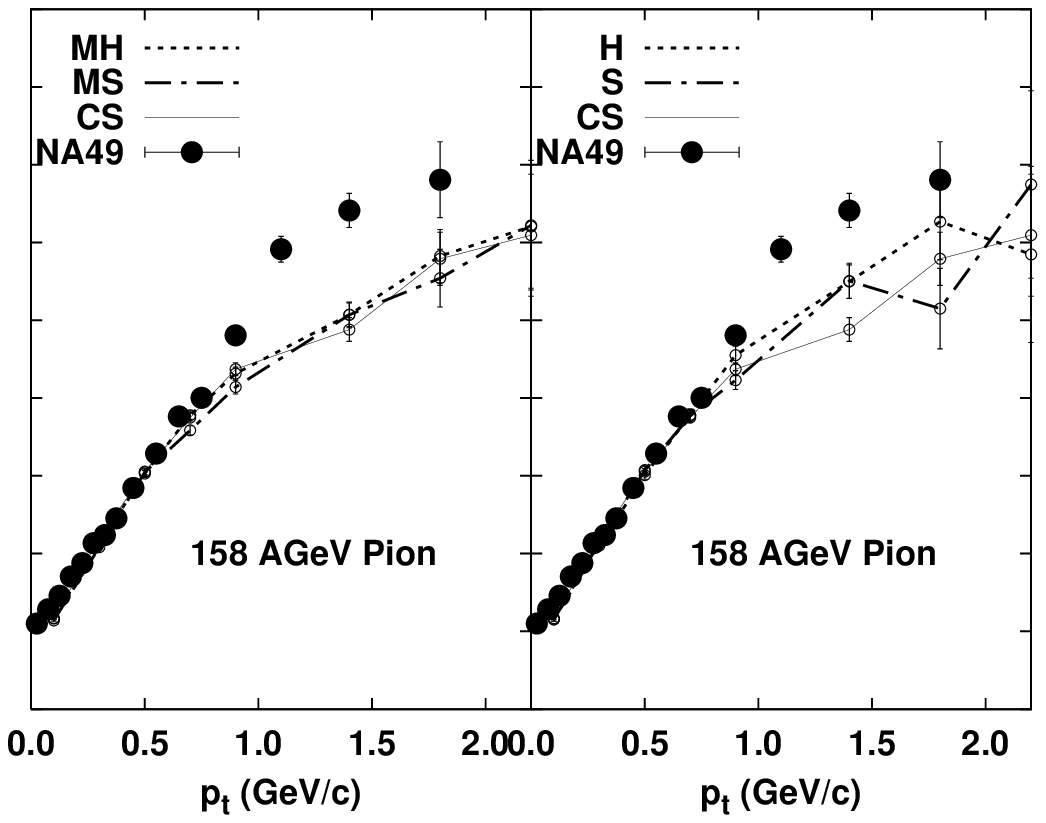}
\caption{
Pion elliptic flow $v_2$ as a function of transverse momentum.
The meaning of the lines is the same as Fig.~\ref{yv2p}.
}
\label{ptv2pi}
\end{figure*}

Rapidity and transverse momentum dependence of the pion $v_2$
are shown in Figs.~\ref{yv2pi} and \ref{ptv2pi}, respectively.
Rapidity dependence at 158$A$ GeV
and transverse momentum dependence at low $p_\ssT$ ($p_\ssT < 1$ GeV/$c$)
at 40$A$ and 158$A$ GeV are well explained by the Cascade model
as well as by the momentum dependent/independent MF models,
and we do not find any significant MF effects for these observables.
By contrast,
we do not see the collapse of $v_2(y)$ at mid-rapidities
seen in the 40$A$ GeV NA49 data,
and we underestimate $v_2$ at high $p_\ssT$.
The former corresponds to the collapse of proton $v_2(y)$ mentioned before.
Momentum independent MF enhances pion $v_2(y)$ slightly,
but this is in the reverse direction to explain the data at 40$A$ GeV.
The strong increase of $v_2(p_\ssT)$ up to around $p_\ssT \sim 2$ GeV/$c$
is also seen at RHIC energies, and this behavior is discussed as
an indication of hydrodynamical evolution~\cite{hydroRHIC}.

%%%%%%%%%%%%%%%%%%%%%%%%%%%%%%%%%%%%%%%%%%%%%%%%%%%%%%%%%%%%%%%%%%%%%%%%%
%  pion v_2 at SPS
%%%%%%%%%%%%%%%%%%%%%%%%%%%%%%%%%%%%%%%%%%%%%%%%%%%%%%%%%%%%%%%%%%%%%%%%%

We now turn to the discussion of the difference between $v_1$ and $v_2$.
We have shown that
MF effects on $v_1$ are rather strong,
but $v_2$ is relatively insensitive to MF at SPS energies.
This may come from the difference of developing time between $v_1$ and $v_2$.
The directed flow $v_1$ at mid-rapidities 
is mainly generated by the interaction
between participants and spectators in the early stage of the collision,
where baryon density is the highest.
On the other hand, $v_2$ in our model is generated in the late stage
until the time reaches the order of nuclear radius,
where densities are not very high,
but it is not formed in the early stage in our model.
This is because our current hadronic transport approach does not
have large participant pressure in the early stages of the collisions,
as we do not explicitly include MF for strings and partonic interactions.
In a hydrodynamic picture, $v_2$ develops from very early times
due to thermal pressure.
This is a striking difference between our approach and hydrodynamics
as previously studied in Ref.~\cite{Sorge1997-99}.

\subsection{Elliptic flow excitation functions from AGS to SPS Energies}

When the incident energy is not high enough,
spectators squeeze participants out of the reaction plane
due to the repulsive nuclear interactions
at 
$0.2\lesssim E_{\rm inc} \lesssim 4A$ GeV.
This squeezing leads to a negative value of the elliptic flow of nucleons
($v_2 < 0$).
The elliptic flow, therefore, 
shows the strength of the repulsive interaction at lower energies.
On the other hand,
elliptic flow becomes positive at higher energies,
because there is no such squeezing effect due to the Lorentz contraction.
Elliptic flow gives a information
how much pressure is generated at higher energies.

\begin{figure*}[thb]
\includegraphics[width=12cm,clip]{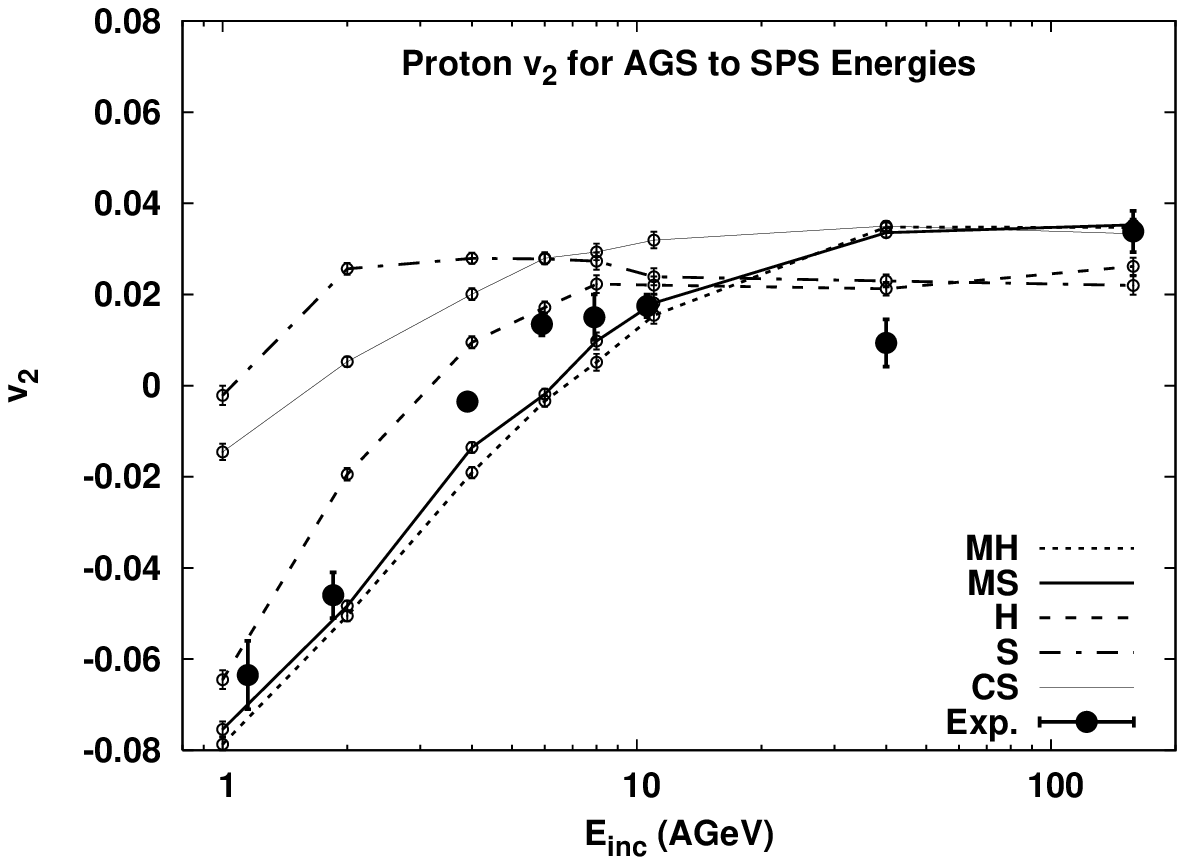}
\caption{Incident energy dependence of proton elliptic flow
at mid-rapidities in mid-central heavy-ion collisions 
from 1$A$ GeV to 158$A$ GeV.
Dotted, bold-solid, dashed, and dot-dashed and thin-solid 
lines show the results of 
Cascade with MH, MS, H, S mean-field,
and Cascade without mean-field (CS),
respectively.
The experimental data are taken from
LBL-EOS, AGS-E895, E877 from Ref.~\cite{E895-99} 
and SPS-NA49 from Ref.~\cite{NA49-03}.
}
\label{sv2}
\end{figure*} 
In Fig.~\ref{sv2},
we show the incident energy dependence of 
proton $v_2$ in mid-central collisions with measured data 
($-0.1<y<0.1$ for AGS, $0<y<2.1 (0<y<1.8)$ for SPS 158$A$ (40$A$) GeV)
\cite{E895-99,NA49-03}.
Rapidity cut $|y| < 0.2 y_{\rm proj}$
has been used in calculations.
Experimental data clearly show the evolution from squeezing
to almond shaped participant dynamics.
With both Cascade and momentum independent
soft
 MF (S),
 we cannot explain strong squeezing effects at lower energies.
The calculated $v_2$ values for momentum independent MF (H,S)
and Cascade are generally larger than data
at AGS energies.
Momentum dependent MF (MH,MS), which is repulsive in the incident energy range
under consideration, pushes down the elliptic flow significantly. 
We qualitatively reproduce the incident energy dependence
from AGS~\cite{E895-99} to SPS~\cite{NA49-03} energies.

Calculated results with both MH and MS are smooth as a function of beam energy,
while the data at $E_\mathrm{inc} = 40A$ GeV has a dip \cite{NA49-03}.
Confirmation of data is necessary
to examine the incident energy dependence of $v_2$,
whether it is a monotonic function or has a dip
at around $E_{\rm inc} \sim 40A$ GeV
by looking at the missing data points.

In our results with momentum dependent potentials,
the stiffness dependence of $v_2$ is smaller
than that in the Boltzmann Equation Model ({\sc bem})~\cite{E895-99,da00,da02}.
In the RQMD/S framework with the relativistic distance $\tilde{\bm r}_{ij}^2$,
the interaction between the projectile and target nucleons are suppressed
at high energies by the factors $m_i/p^0_i$ and $m_j/p^0_j$
in the potential derivatives in Eqs.~(\ref{Eq:Dij}) and (\ref{Eq:Eij}).
For momentum dependent potentials, 
we have the relative momentum vector ${\bm p}_{ij}$ in Eq.~(\ref{Eq:dpsdp}),
which can compensate the suppression factor in Eq.~(\ref{Eq:Eij}).
For momentum independent potentials, on the other hand,
the pair velocity $\bm\beta_{ij}$ in Eq.~(\ref{Eq:PairGamma}) is very small
for nucleon pairs between the projectile and target,
and there is no enhancement factor to compensate the above suppression
in the derivatives of the relativistic distance $\tilde{\bm r}_{ij}^2$ 
in Eqs.~(\ref{Eq:drsdp}) and (\ref{Eq:drsdr}).
This suppression does not happen in {\sc bem},
and they find significant stiffness dependence in Refs.~\cite{E895-99,da00,da02},
while we do not see strong stiffness dependence.
In the case of momentum independent potentials,
our results are closer to the Cascade results
compared to those in Ref.~\cite{urqmd2,urqmd3}.
This difference also comes from 
the above suppression between the projectile and target nucleons.
Essential reason for those difference is that in {\sc rqmd} or RQMD/S, 
potentials are regarded as Lorentz scalar.
Possible other model dependence will be discussed in the next section.

\section{Model Uncertainties}
\label{sec:discussion}

In the previous section,
it has been shown that
the momentum dependent hard or soft MF
improved the description of the collective flow data
from AGS to SPS energies.
However, there are some uncertainties in our calculations
for the study of collective flows.

\begin{figure*}[thb]
\begin{center}
\includegraphics[width=9.5cm,clip]{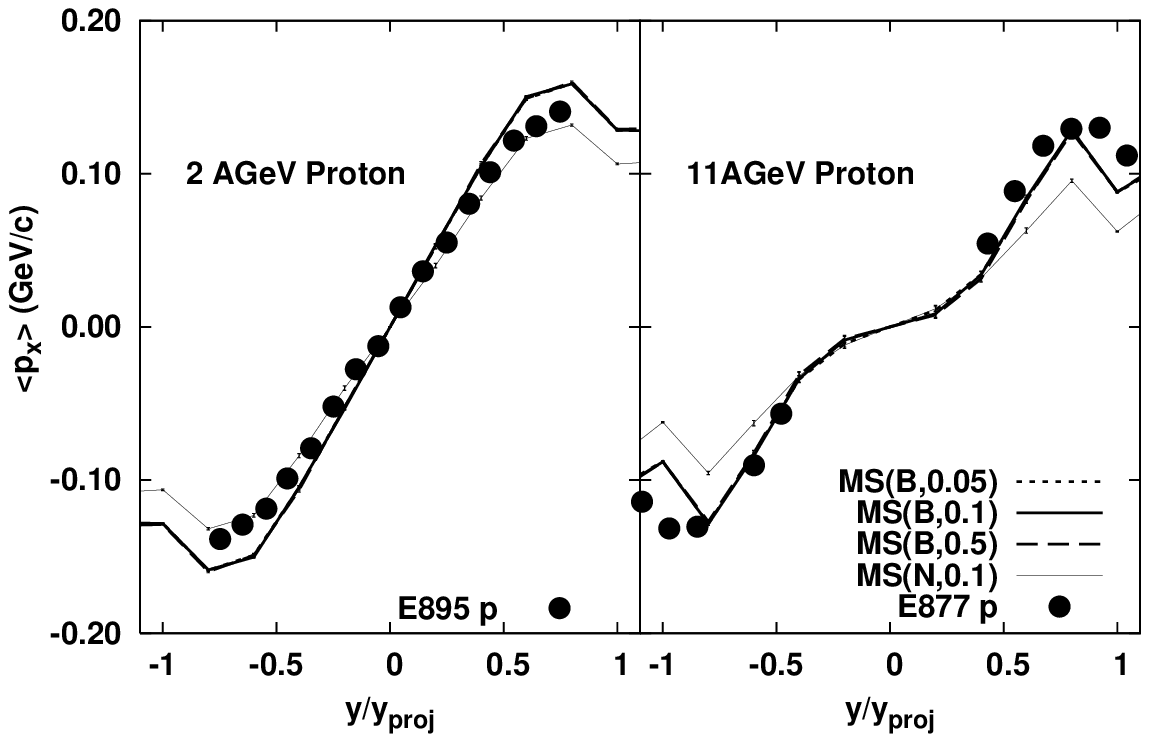}~\hspace{-0.5cm}
\includegraphics[width=8.5cm,clip]{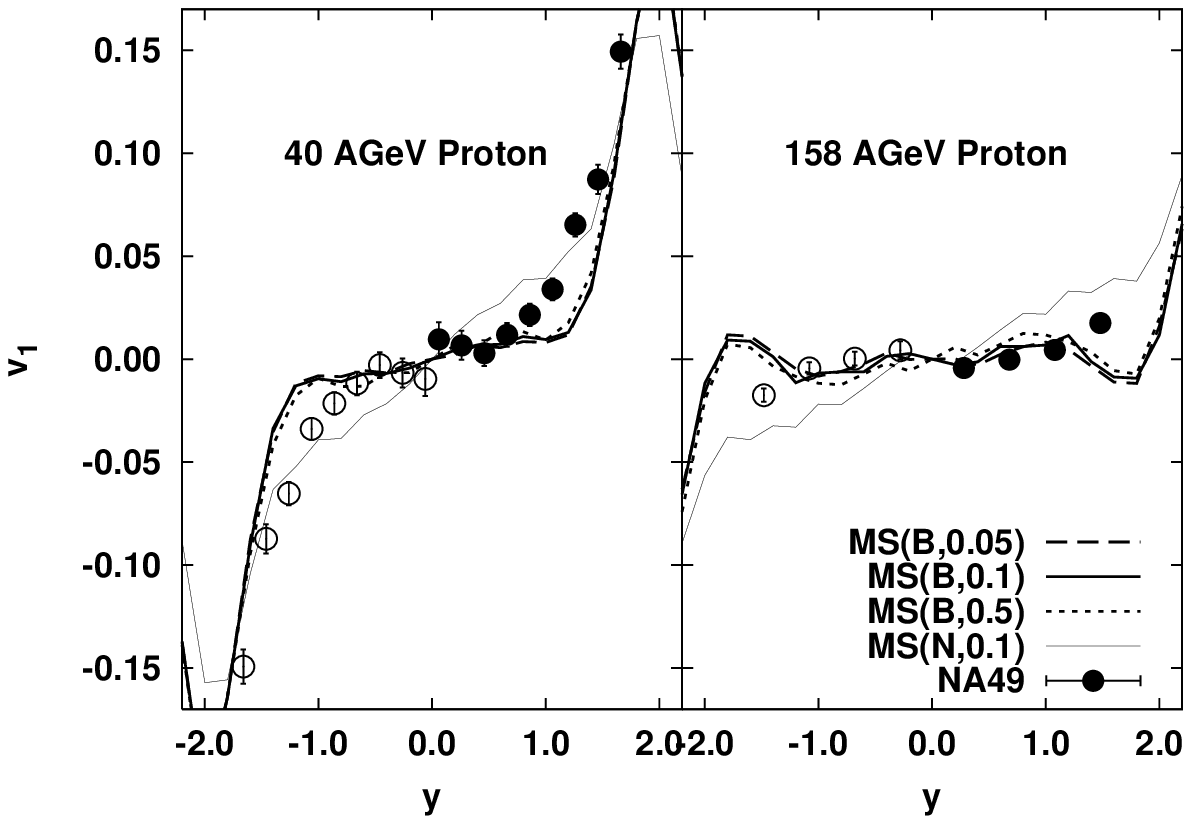}
\end{center}
\caption{
The sideward (directed) flows
of AGS (left) and SPS (right) energies
are compared with different
assumptions for the mean-fields.
Momentum dependent soft (MS) type mean-field 
has been used in these calculations.
The first term in the parenthesis means that
all baryon (B) or only nucleon (N) are included
for the MF.}
The second term (0.05/0.1/0.5) denotes simulation time step size $dt$
(fm/$c$).
\label{ypxb11}
\end{figure*}

\begin{figure*}[thb]
\begin{center}
\includegraphics[width=10cm,clip]{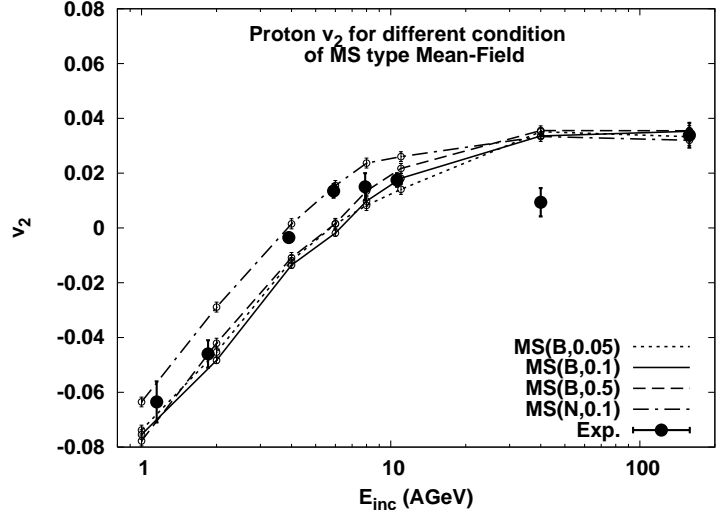}
\end{center}
\caption{
The elliptic flows
are compared with different
assumptions for the mean-fields.
Meaning of the lines is the same as Fig.~\ref{ypxb11}.
}
\label{sv2b}
\end{figure*}

% MF ``B''
First, let us consider the effects of the MF for non-nucleonic baryons.
Strange baryons, resonance hadrons or
anti-baryons are expected to feel MF,
which may be different from that for nucleons.
In the previous section, we have assumed that all the baryons
feel the same MF,
and this treatment would give a rough estimate of a maximum MF effects,
since, for example,
MF for $\Delta$'s or $\Lambda$'s is generally expected to
be smaller than that of nucleons.
On the other hand, 
if we include MF only for nucleons,
we may get a rough estimate of a minium baryonic MF effects,

In Figs.~\ref{ypxb11} and \ref{sv2b},
we compare the results with and without MF for
non-nucleonic baryons.
We can read in the left panel of 
Fig.~\ref{ypxb11} that 
ignoring MF for non-nucleonic baryons
(specified as ``N'' in the figure) at 2A GeV 
reduces both the $\langle p_x\rangle$ slope
and the strength  at the forward rapidities by about 20 \%
compared to the case of ``B'' in which
all baryons feel MF.
But slope remains the same at 11A GeV among ``N'' and ``B''.
On the other hand, 
it is seen in the right panel of Fig.~\ref{ypxb11}
that MF only for nucleons
is not enough to suppress $v_1$ at SPS energies.
One can also see some differences of $v_2$ in ``B'' and ``N'' 
in Fig.~\ref{sv2b} up to AGS energies.
The experimental data lie between
``B'' and ``N'' except for 40 $A$ GeV,
suggesting that MF for excited baryons are smaller than that
of nucleons.

Next, we have checked the time step size $dt$ dependence.
Since update of MF after each collision requires a huge calculation time,
we evaluate MF only at each time
slice.
When a baryon collides in one time step,
that baryon is propagated with MF until the collision time 
and we ignore the MF after the collision before it is formed.
In the time step of baryon formation,
displacements by the MF for ${\bm p}_i$ and ${\bm r}_i$ are evaluated
by using the MF at the next time slice.
This treatment is valid up to the first order in $dt$
when one baryon collides and/or is formed once in one time step.
In the later stages,
this prescription is expected to work well
because of the low collision frequency.
In the early stages, many collisions make strings and resonances
which do not feel MF, so our prescription may not be too bad.
In the middle stages, however, it may be possible that
elastic scatterings are frequent enough and baryons
keep to feel MF after collisions in each time step.
Thus we need to analyze the collision frequency effects
on constructing flows by reducing $dt$.
In Figs.~\ref{ypxb11} and \ref{sv2b}, we plot the results
with different time step sizes $dt$=0.05, 0.1 and 0.5 fm/$c$.
For $\langle p_x\rangle$ at AGS energies (left panel of Fig.~\ref{ypxb11}), 
all the results with different time step size agree well with each other, 
and we cannot distinguish these lines.
Time step size dependence of the $v_2$ as shown in Fig.~\ref{sv2b}
still gives us the confirmation of the convergence of the numerical results.
For $v_1$ at SPS energies (right panel of Fig.~\ref{ypxb11}),
only very small differences
can be seen between the results with $dt=0.5$ fm/$c$ and $dt\leq 0.1$ fm/$c$.
We conclude that $dt=0.1$ fm/$c$ which has been used
as a default throughout this work
is small enough to perform a reliable calculations.

%%%%%%%%%%%%%%%%%%%%%%%%%%%%%%%%%%%%%%%%%%%%%%%%%%%%%%%%%%%%%%%%%%%%%%%%%%%%%%%%
Finally, we would like to address the problem of
the uncertainties of the transport model itself.
In addition to the ambiguities in introducing collision terms,
the equations of motion depend on the model treatment.
It is not trivial at all
to construct equations of motion of relativistic particles
during heavy-ion collisions
based on the potential or the MF
giving an appropriate EOS.
At relativistic energies, there are several ways proposed so far
to introduce the potential effects.
\begin{enumerate}
\item Relativistic Mean Field (RMF) having Lorentz
scalar $U_s$ and vector $U_v^\mu$ terms ({\sc rbuu}~\cite{sahu2,Maruyama1993}).
The scalar and the vector time-component are evaluated
in the local rest frame, and by the Lorentz transformation 
we can get $U_v^\mu$ in the calculation frame.
Thus neglecting a nonlocality in time,
this evaluation of the potential is practically covariant.
In this approach, however,
we need to introduce strong cut-off for the coupling of
vector meson and baryons~\cite{sahu2},
since the vector potential effects linearly increase as a function
of incident energy.
\item Lorentz scalar re-interpretation of non-relativistic
potentials ({\sc buu}~\cite{Giessen}, {\sc bem}~\cite{da00}).
In the {\sc buu} model~\cite{Giessen},
the Lorentz scalar MF $U_s$ is obtained
from the non-relativistic MF $U$ in the local rest frame
through the relation,
\begin{align}
\varepsilon(\bm{p},\rho)
&=\sqrt{(m + U_s(\bm{p},\rho))^2 + \bm{p}^2}
\nonumber\\
&= \sqrt{m^2 + \bm{p}^2} + U(\bm{p},\rho)\ ,
\end{align}
where $\rho$ is the baryon density~\cite{Giessen}.
For the momentum independent MF in {\sc bem} in Refs.~\cite{da00,da02},
the scalar potential is directly given
so as to fit the EOS, and the scalar density is used for $\rho$.
They do not have any vector terms increasing at high energies,
and the potential effects become mild compared to the RMF treatment.
For example, the derivative of the above single particle energy
gives rise to the factor $(m+U_s)/\varepsilon$ in front of
the $U_s$ derivative, and suppresses the potential effects. 
\item Combination of the Lorentz scalar
and non-relativistic type density dependent potentials
({\sc bem}~\cite{da98,da00,da02}).
This approach is adopted in Ref.~\cite{da98}
and in the momentum dependent MF in Refs.~\cite{da00,da02}.
The single particle energy is given as
\begin{align}
&\varepsilon(p,\rho)=m+\int_0^p dp' v^*(p' ,\rho) + \widetilde{U}(\rho)\ ,
\\
&v^*(p,\rho)={p \over \sqrt{p^2 + \left(m^*(p,\rho)\right)^2}}\ .
\end{align}
The derivative of $\widetilde{U}(\rho)$
does not come with the suppression factor such as $m/\varepsilon$,
and generates strong effects at high energies,
where both of the density and density derivative become large.
\item	Constraint Hamiltonian dynamics
({\sc rqmd}~\cite{sorge}, RQMD/S~\cite{RQMD_S}).
In {\sc rqmd} and RQMD/S, particle velocity and force are not given
by the derivatives of the single particle energy,
but by the derivatives of the total Hamiltonian,
as shown in the Appendix in the case of RQMD/S.
Thus the relation to other MF models described above is not
straightforward.
However, the potential $V_i$
in the on-mass-shell constraint (\ref{Eq:On-Mass-Shell})
is introduced as Lorentz scalar,
and we have suppression factor $m/p^0$ in the equations of motion,
(\ref{Eq:EOM-R}) and (\ref{Eq:EOM-P}).
These observations suggest that the {\sc rqmd} and RQMD/S would give results
similar to those in Lorentz scalar MF models,
such as {\sc buu}~\cite{Giessen}.
Another difference from other MF models
exists in the nuclear density profile.
One nucleon is represented by one Gaussian packet
rather than many test particles,
then the nuclear diffuseness becomes generally larger in QMD-type models.
This may generate artificial surface effects at large impact parameters
or in light-ion collisions.
However, central and mid-central collisions of heavy-nuclei 
are expected to be well described,
as in the case of various cascade models,
in which one particle is used for one hadron.
\end{enumerate}
We are not very sure which is the best way
to include the potential effects in high-energy heavy-ion collisions.
Further formal developments on transport models in relation to nuclear EOS
would be necessary,
and at the same time, phenomenological studies of heavy-ion collisions
are required to verify the validity of models and to elucidate the EOS.
From the latter point of view, systematic study in a wide range of
incident energy is needed, since the above uncertainties are 
closely related to the Lorentz transformation properties,
whose effects would vary drastically as the incident energy varies.
The incident energy range from AGS to SPS energies studied in this work
may provide a good benchmark test for transport models and the EOS.

\section{Summary}
\label{sec:summary}
\indent
We have investigated collective flows in heavy-ion collisions
from  AGS ($(2-11)A$ GeV) to SPS ($(40,158)A$ GeV) energies
by using a combined framework of hadron-string cascade ({\sc jam})~\cite{jam}
and covariant constraint Hamiltonian dynamics (RQMD/S)~\cite{RQMD_S}.
In {\sc jam}, various particle production mechanisms
are taken into account --- production and decay of resonances and strings,
jet production and its fragmentation.
Momentum dependence of the MF is fitted~\cite{maru2} to the real part
of the Schr{\"o}dinger equivalent global optical potential
of Hama et al.~\cite{Hama1990}
in a Lorentzian form Eq.(\ref{usep}).
Saturation properties are fitted
by introducing the density dependent
potential of Skyrme-type in the power series of $\rho$.
Calculated results of Cascade,
Cascade with momentum dependent MF,
and Cascade with momentum independent MF
are compared with the data of
sideward $\langle{p_x}\rangle$,
directed $v_1$,
and elliptic $v_2$ flows
as a function of rapidity, transverse momentum and beam energy
from AGS to SPS.
Generally, results with momentum dependent MF 
reasonably well explain the trend of the data for proton flows.
We note that it is for the first time to explain
anisotropic proton collective flow data of heavy-ion collisions
from AGS to SPS in one framework consistently. 
Without momentum dependence in MF,
we cannot reproduce the strong enhancement of the sideward flow
at $E_{\rm inc} = (2-11)A$ GeV,
strong squeezing seen in $v_2$ for $E_{\rm inc} \lesssim 4A$ GeV, 
and 
the suppression of proton $v_1$ at $E_{\rm inc} = 40A$ and $158A$ GeV.

Our new model ---
hadron-string cascade with momentum dependent MF ---
provides an improved description for collective flows
in mid-central collisions from AGS to SPS energies.
The present analysis implies that the effects of the
momentum dependent potential
is large up to the SPS energies.

There are still many problems to pin down the equation of state
of dense nuclear matter from heavy-ion data.
First, we have made an assumption that
MF is taken into account only for baryons
and all the baryons feel the same MF.
It is interesting to extend the present work to
discuss the MF effects for mesons
and different MF for hyperons and resonance hadrons,
and look at the $\Lambda$ or kaon flow data.
Secondly,
we cannot make soft EOS ($K \sim 200$ MeV) in the present form of MF
which is consistent with the optical potential.
When the momentum dependence is fitted to the optical potential
by Hama et al.~\cite{Hama1990} in the Lorentzian form,
the EOS necessarily becomes relatively stiff
in combination with the Skyrme-type density dependent form
as shown in Table~\ref{parsets}.
The small sensitivity on EOS with momentum dependence appeared in this work
may be suggesting that the probed EOS range is not wide enough.
Finally, the model dependence of the MF treatment has to be cleared
in order to obtain model independent EOS information.
For this purpose, it is necessary to test various MF treatment
in one framework.
In the present model,
we need to modify the on-mass-shell constraint
to include Lorentz vector potentials or potentials of other types.
It can be a breakthrough
for a transport model-independent discussion of EOS.

\acknowledgements

We are grateful to Professor Tomoyuki Maruyama
and Professor Pawel Danielewicz for useful discussions and comments.
This work is supported in part by the Ministry of Education,
Science, Sports and Culture,
Grand-in-Aid for Scientific Research (C)(2), No. 15540243, 2003.

\appendix

\section{RQMD/S formalism}
Here we briefly summarize the RQMD/S formalism
developed by Maruyama et al. in Ref.~\cite{RQMD_S} for completeness.
The original RQMD formalism is initiated by Sorge et al. in
Ref.~\cite{sorge}.
RQMD(/S) is based on
the constrained Hamiltonian dynamics~\cite{komar}
which is formulated in a manifestly covariant way.
We use 4-vectors $q^\mu_i$ and $p^\mu_i$ for the description of
the $N$ particle system.
Therefore, we need to have $2N$ constrains $\phi_i  (i=1,\dots,2N)$
as physical phase space is $6N$ dimension.
Now our Hamiltonian may be constructed from the constraints $\phi_i$
and Lagrange multiplier $u_i$
from the Dirac's constraint Hamiltonian formalism
\begin{align}
 H = \sum_{i=1}^{2N-1}u_i \phi_i \label{uphi}.
\end{align}
The equations of motion are then
\begin{align}
&\mbox{\hspace{-0.2cm}} 
\frac{dq_i}{d\tau} = \{H,q_i\}
\approx\sum_{j=1}^{2N-1}u_j\frac{\partial \phi_j}{\partial p_i},
\label{dq1} \\
&\mbox{\hspace{-0.2cm}} 
\frac{dp_i}{d\tau} = \{H,p_i\}
\approx -\sum_{j=1}^{2N-1}u_j\frac{\partial \phi_j}{\partial q_i}
\label{dp1},
\end{align}
where the Poisson brackets are defined as
\begin{align}
 \{A , B\}  \equiv \sum_{k,\mu} \Biggl( 
 \frac{\partial A}{\partial q_k^\mu} &\frac{\partial B}
{\partial p_{k \mu}} -
 \frac{\partial A}{\partial p_{k \mu}} \frac{\partial B}{\partial q_k^\mu}
 \Biggr), \\
 \{ q_{i}^{\mu} , p_{j \nu} \} = \delta^{\mu}_{\nu} \delta_{i j},\quad
 \{ q_{i}^{\mu} , q_{j}^{\nu} \}& =  0 ,\quad  
 \{ p_{i \mu} , p_{j \nu} \} = 0, \\
 i,j,k = 1,\dots,N & \notag \; , \;\;\mu,\nu=0,1,2,3 
\end{align}
and the sign ``$\approx$'' means the weak equality
initiated by Dirac~\cite{dirac}.
When we require that 
constraints $\phi_i$ should conserve
in time, then they fulfill
\begin{align}
 \frac{d\phi_i}{d\tau}=\frac{\partial\phi_i}{\partial\tau} +
 \sum^{2N-1}_{j=1}u_j\{ \phi_i , \phi_j\} \approx 0.
\end{align}
Particle trajectories in $6N$ phase space is uniquely determined by the
equations of motion Eqs.~(\ref{dq1}) and (\ref{dp1})
together with Eqs.~(\ref{ui}) and (\ref{cij})
when $2N$ constraints are given.
 
We use the following $2N$ constraints in RQMD/S
\begin{equation}
 \phi_i \equiv \left\{
\begin{array}{ll}
H_i \; ,\quad &i=1,\dots,N \\
\chi_{i-N}\; ,\quad& i=N+1,\dots,2N. 
\end{array}
\right.
\end{equation}
First $N$ constraints are the on-mass-shell constraints
\begin{align}
\label{Eq:On-Mass-Shell}
 H_i \equiv p_i^2 - m_i^2 - 2m_iV_i \approx 0\, ,\quad i= 1,\dots,N .
\end{align}
Remaining $N$ conditions constrain the time fixation of the particles.
In original RQMD time fixation~\cite{sorge,sorge2,maru},
$N\times N$-matrix has to be solved numerically at each time step
to deduce inverse matrices.
Moreover, if particle production or annihilation occurs,
the time fixation is violated and
a initial $q_i$ of produced particles
satisfying the constraints and energy conservation
has to be imposed.

Maruyama et al. introduced  simplified
time fixation in RQMD/S with global time parameter $\tau$ in Ref.~\cite{RQMD_S}
as
\begin{equation}
\begin{array}{ll}
 \chi_i \equiv \hat{a}\cdot (q_i - q_N) \approx 0  \, ,\quad i = 1,\dots,N-1,\\
 \chi_N \equiv \hat{a}\cdot q_N - \tau \approx 0 , 
\label{chiiN}
\end{array}
\end{equation}
where, $\hat{a}$ is a 4-component vector corresponding to
$(1,{\bm 0})$ at the rest frame of the particle,
and $q_i$ is space-time coordinate of the $i$-th particle.
The constraints Eq.~(\ref{chiiN}) 
is able to be kept in the case of particle productions.

Since the constraints $\phi_i (i=1,\dots,2N-1)$ does not depend explicitly on $\tau$,
Lagrange multiplier $u_i(\tau)$ can be solved as
\begin{align}
& u_i \approx -\frac{\partial \phi_{2N}}{\partial\tau}C_{2N,i} ,
\quad i=1,\dots,2N-1 \label{ui},\\
&\mbox{where}\quad C^{-1}_{ij} \equiv \{\phi_i,\phi_j\} ,\quad i,j=1,\dots,2N
\label{cij}.
\end{align}
The matrix $C$ (inverse of matrix $C^{-1}$) must exist,
because we only allow for the  $\tau$-dependent $2N$th constraint
functions which combine with $2N-1$ constraints. 
Furthermore,
$C$ can be obtained analytically,
if we replace $p^0_i$ in the potential $V_i$
with the kinetic energy $\sqrt{\bm p^2_i+m^2_i}$.
This is a great advantage in the point of view of CPU time.
One obtains the RQMD/S Hamiltonian
\begin{equation}
 H \approx\sum_{i=1}^{N} u_i (p_i^2-m_i^2-2m_iV_i)\ ,
\label{Hrel}
\end{equation}
where
\begin{equation}
u_i=\frac{1}{2p_i^0},\quad p_i^0=\sqrt{\bm p_i^2 + m_i^2 + 2m_iV_i}\ .
\label{eq13}
\end{equation}
The equations of motion are then
\begin{align}
\label{Eq:EOM-R}
\frac{d\bm r_i}{d\tau}&\approx
-\frac{\partial H}{\partial \bm p_i}
=\frac{\bm p_i}{p_i^0}+\sum_{j=1}^{N}\frac{m_j}{p_j^0}
\frac{\partial V_j}{\partial\bm p_i},\\
\label{Eq:EOM-P}
\frac{d\bm p_i}{d\tau}&\approx
\frac{\partial H}{\partial \bm r_i}
= -\sum_{j=1}^{N}\frac{m_j}{p_j^0}
\frac{\partial V_j}{\partial\bm r_i} .
\end{align}
In actual calculations,
we have replaced $p^0_i$ with the kinetic energy $\sqrt{\bm p^2_i+m^2_i}$
in the denominators of Eqs.~(\ref{Eq:EOM-R}) and (\ref{Eq:EOM-P})
after evaluating all the derivative terms
for simplicity.
This approximation would be valid in the relativistic energy region,
where the kinetic energy is much larger than the potential $V_i$.

Relative distance $\bm r_{ij}= \bm r_i - \bm r_j$
and 
$\bm p_{ij}= \bm p_i - \bm p_j$
in the potentials
should be replaced by
the squared 4-vector distance with Lorentz scalar as,
\begin{align}
- q_{T ij}{}^{2} = - q_{ij}{}^{2} + 
\frac{(q_{ij}\cdot P_{ij})^{2}}{P_{ij}{}^{2}}, \label{qtij}\\
- p_{T ij}{}^{2} = - p_{ij}{}^{2} + 
\frac{(p_{ij}\cdot P_{ij})^{2}}{P_{ij}{}^{2}}, \label{ptij}
\end{align}
where
$p_{ij} = p_{i}-p_{j}\; ,
q_{ij} = q_{i}-q_{j}\; ,\quad P_{ij} =  p_{i} +p_{j}$.
We note that in non-relativistic limit,
$-q_{T ij}{}^2\xrightarrow[c\to\infty]{}\bm r_{ij}^2$.
This assumption takes into account
the contraction of longitudinal direction and
we can avoid unphysical compression.
In the actual simulations, we use the following expression
\begin{eqnarray}
-q_{Tij}^2 &\equiv&
 \tilde{\bm r}_{ij}^2
=\bm r_{ij}^2 
+ \gamma_{ij}^2(\bm r_{ij}\cdot \bm\beta_{ij})^2, \\
- p^2_{T ij} &\equiv&
 \tilde{\bm p}_{ij}^2 = \bm p_{ij}^2 
      -(p_i^0-p_j^0)^2+ \gamma_{ij}^2\biggl(
      \frac{m_i^2-m_j^2}{p_i^0+p_j^0}
\biggr)^2\ ,
\nonumber\\
\end{eqnarray}
where, the velocity and $\gamma$-factor between $i$- and $j$-th
particle are given by
\begin{equation}
\label{Eq:PairGamma}
\bm\beta_{ij}= \frac{\bm p_i+\bm p_j}{p_i^0+p_j^0}\ ,
\quad
\gamma_{ij}=\frac{1}{\sqrt{1-\bm\beta_{ij}^2}}\ .
\end{equation}

We now write down the explicit form of the equations
of motion in RQMD/S which is used in the actual simulation.
As explained in Sec.~\ref{sec:eos},
we use the following potentials
\begin{eqnarray}
V &=& \sum_i ( V_{\rm Sky\; \it i}+ V_{\rm mom \;\it i}) \nonumber\\
 &=& \sum_i\biggl[
  \frac{\alpha}{2\rho_0}
\langle\rho_i\rangle
  + \frac{\beta}{(1+\gamma)\rho_0^\gamma}
\langle\rho_i\rangle
^{\gamma} \nonumber\\
&+& 
\sum_{k=1,2}
\frac{C_{\rm ex}^{(k)}}{2\rho_0}\sum_{j(\ne i)}\frac{1}{1+[\tilde{\bm
 p}_{ij}/\mu_k]^2}\rho_{ij}
\biggr],
\end{eqnarray}
where $\langle\rho_i\rangle$ is obtained from a convolution of
Gaussian wave packet:
\begin{align}
 \langle \rho_i \rangle &
\equiv\sum_{j(\ne i)}\int d\bm r \rho_i(\bm r)\rho_j(\bm r)
=\sum_{j(\ne i)}\rho_{ij} \nonumber\\
&=\sum_{j_(\ne i)}
\frac{1}{(4\pi L)^{\frac32}}\exp\biggl(
-\frac{\tilde{\bm r}_{ij}^2}{4L}\biggr).
\label{Eq:rhoi}
\end{align}
The width parameters $L=$2.05(MH), 2.1(MS), 1.08(H and S) fm$^2$
are taken from Refs.~\cite{maru2,aich}.
The equations of motion
Eqs.~(\ref{Eq:EOM-R}) and (\ref{Eq:EOM-P}) then become
\begin{align}
\label{Eq:drdtau}
\frac{d\bm r_i}{d\tau} &
=\frac{\bm p_i}{p_i^0}+
\sum_{j(\ne i)}D_{ij}
 \frac{\partial\tilde{\bm r}^2_{ij}}{\partial\bm p_i}
+ \sum_{j(\ne i)}E_{ij}
\frac{\partial\tilde{\bm p}^2_{ij}}{\partial\bm p_i} \\
\label{Eq:dpdtau}
\frac{d\bm p_i}{d\tau} &
= -\sum_{j(\ne i)} D_{ij}
\frac{\partial\tilde{\bm r}^2_{ij}}{\partial\bm r_i}
\end{align}
where,
\begin{align}
\label{Eq:Dij}
 D_{ij}&=
\biggl(-\frac{1}{2L}\biggr)\rho_{ij} 
\left[ 
 \frac{\alpha}{2\rho_0}\biggl(\frac{m_i}{p_i^0}+\frac{m_j}{p_j^0}\biggr) \right.
 \nonumber\\
&\left.
+ \frac{\gamma}{1 + \gamma}
\frac{\beta}{\rho_0^\gamma}
\Biggl\{ \frac{m_i}{p_i^0}
\langle\rho_i\rangle
^{\gamma -1}
\!\!+ \frac{m_j}{p_j^0}
\langle\rho_j\rangle
^{\gamma -1}
\Biggr\}
\right]\nonumber \\
&+
\biggl(-\frac{1}{4L}\biggr)
\frac{1}{2\rho_0}
\rho_{ij}
\biggl(\frac{m_i}{p_i^0}+\frac{m_j}{p_j^0}\biggr) 
\sum_{k=1,2}
\frac{C_{\rm ex}^{(k)}}{1+[\tilde{\bm
 p}_{ij}/\mu_k]^2},\\
\label{Eq:Eij}
E_{ij}&=
\frac{1}{2\rho_0}
\rho_{ij}
\biggl(\frac{m_i}{p_i^0}+\frac{m_j}{p_j^0}\biggr) 
\sum_{k=1,2}
\biggl(-\frac{1}{\mu_k{}^2}\biggr)
\frac{C_{\rm ex}^{(k)}}{1+[\tilde{\bm
 p}_{ij}/\mu_k]^2}
.
\end{align}
The result of the differentials are~\cite{hirataphd}
\begin{align} 
\label{Eq:drsdp}
\frac{\partial\tilde{\bm r}^2_{ij}}{\partial\bm p_i}
&=\frac{2\gamma_{ij}^2}{p_i^0+p_j^0}(\bm r_{ij}\cdot\bm \beta_{ij})
\biggl\{ 
\bm r_{ij}+ \gamma_{ij}^2(\bm r_{ij}\cdot\bm\beta_{ij})
\biggl(\bm\beta_{ij}-\frac{\bm p_i}{p_i^0}\biggr) \biggr\}\ , \\
\label{Eq:drsdr}
\frac{\partial\tilde{\bm r}^2_{ij}}{\partial\bm r_i}
&= 2\bm r_{ij} + 2\gamma_{ij}^2(\bm r_{ij}\cdot\bm\beta_{ij})
\bm\beta_{ij} \ , \\
\label{Eq:dpsdp}
\frac{\partial\tilde{\bm p}^2_{ij}}{\partial\bm p_i}
&= 2\bm p_{ij}-2(p_i^0-p_j^0)\frac{\bm p_i}{p_i^0}
\nonumber\\
&\quad
+2\gamma_{ij}^4 \frac{1}{p_i^0+p_j^0}
\biggl(\frac{m_i^2-m_j^2}{p_i^0+p_j^0}\biggr)^2
\biggl(\bm \beta_{ij}-\frac{\bm p_i}{p_i^0}\biggr).
\end{align}

Finally, let us check the non-relativistic limit to confirm
the validity of Eqs.~(\ref{Hrel})--(\ref{Eq:EOM-P}).
We define kinetic energy as $\mathcal E_i \equiv p_i^0 - m_ic^2$
(here we write the speed of light $c$ explicitly),
Indeed Hamiltonian Eq.~(\ref{Hrel}) and the equations of motion
have the correct non-relativistic limit as
\begin{align}
 H \approx \sum_{j=1}^{N}&\frac 12 \frac{1}{\mathcal E_j / c^2 + m_j}
\biggl(\frac{\mathcal E_j^2}{c^2}+2m_j\mathcal E_j - \bm p_j^2 
  -2m_jV_j \biggr) \notag\\
&\mbox{\hspace{-1.1cm}} 
\mathop{\approx}_{c\to\infty}
 \sum_{j=1}^{N}\biggl( \mathcal E_j - \frac{\bm p_j^2}{2m_j}
  - V_j
\biggr)
= E - H_\mathrm{N. R.}\ ,
\\
 \frac{d\bm r_i}{d\tau}&=\frac{\partial H_\mathrm{N. R.}}{\partial \bm p_i}
\approx
\frac{\bm p_i}{m_i}+\sum_{j=1}^{N}
\frac{\partial V_j}{\partial\bm p_i}\ ,\\
\frac{d\bm p_i}{d\tau}&=-\frac{\partial H_\mathrm{N. R.}}{\partial \bm r_i}
\approx -\sum_{j=1}^{N}
\frac{\partial V_j}{\partial \bm r_i} \ .
\end{align}

\end{document}